\RequirePackage{fix-cm}

\documentclass[twocolumn,epjc3]{svjour3}  
\smartqed  
\RequirePackage{mathptmx}      
\RequirePackage{graphicx}
\RequirePackage{amssymb}
\RequirePackage{multirow}
\RequirePackage{amsmath}
\RequirePackage{lineno}
\RequirePackage{booktabs}
\RequirePackage{float}
\RequirePackage{comment}
\RequirePackage[version=4]{mhchem}
\RequirePackage{booktabs}
\RequirePackage{url}
\RequirePackage[colorlinks,allcolors=red]{hyperref}
\RequirePackage[multiple]{footmisc}
\RequirePackage{upgreek}
\RequirePackage{mathtools}
\RequirePackage{cuted}
\RequirePackage{relsize}
\RequirePackage{makecell}
\RequirePackage{tablefootnote}
\RequirePackage{siunitx}

\journalname{Eur. Phys. J. C}

\makeatletter
\renewcommand{\thanksref}[1]{\nolinebreak\textsuperscript{\ref{#1}}\nolinebreak\checknextarg}
\newcommand{\checknextarg}{\@ifnextchar\bgroup{\nolinebreak\gobblenextarg}{}}
\newcommand{\gobblenextarg}[1]{ \textsuperscript{\nolinebreak\hspace{-4pt}\mbox{\nolinebreak$^,$\nolinebreak\ref{#1}\nolinebreak}\nolinebreak} \@ifnextchar\bgroup{\gobblenextarg}{}}

\begin{document}

\title{Cryogenic characterisation of the FBK NUV-HD-cryo Silicon Photomultipliers for the DarkSide-20k dark matter experiment}
 \widowpenalty10000
  \clubpenalty10000

\author{The DarkSide-20k Collaboration$^\text{\normalfont a,1}$}
\thankstext{e1}{\vspace{1cm}corresponding authors: ds-ed@lists.infn.it}
\institute{See back for author list \label{addr1}}

\date{Received: date / Accepted: date}

\maketitle

\modulolinenumbers[5]

\begin{abstract}
Liquid Argon Time Projection Chambers (TPCs) are highly promising detectors for dark matter searches due to their uniform response, scalability to large target masses, and compatibility with ultra-low background conditions. The DarkSide-20k (DS-20k) experiment, currently under construction at Istituto Nazionale di Fisica Nucleare (INFN) Laboratori Nazionali del Gran Sasso (LNGS), aims to achieve unprecedented sensitivity to Weakly Interacting Massive Particles (WIMPs), approaching the neutrino fog. At the heart of the detector is a dual-phase TPC filled and surrounded by low-radioactivity argon from underground sources ({\it i.e.} Underground Argon (UAr)), functioning both as the WIMP target and detection medium. The experiment employs Fondazione Bruno Kessler (FBK) NUV-HD-cryo Silicon Photomultipliers (SiPMs) as the chosen photon sensors for its optical planes. 
Key SiPM properties such as dark noise rate, gain and direct crosstalk were measured using specialized test setups at \SI{77}{\kelvin}. The photon detection efficiency (PDE) was measured under vacuum at 77~K over the wavelength range 350–820~nm. Secondary photon emission, which can give rise to external optical crosstalk, was measured under vacuum at 86~K over the wavelength range 550–1020 nm.
\end{abstract}

\section{Introduction}
\label{sec:intro}

Silicon photomultipliers (SiPMs) have emerged as a compelling photo-sensor solution for detecting single photons in various applications, ranging from particle physics to medical imaging~\cite{Garutti2011}. SiPMs consist of an array of tightly packed Single Photon Avalanche Diodes (SPADs) with quenching resistors, operated above the breakdown voltage, to generate self-sustaining charge avalanches upon absorbing an incident photon.   Photomultiplier tubes (PMTs) are widely used for light detection in liquid-xenon experiments, and dedicated development efforts have substantially reduced their intrinsic radioactivity~\cite{Abe2019,Aprile2015}. SiPMs provide an alternative whose radiopurity depends on the sensors, the substrate material, and electronics used in the photodetector assembly. For example, a fully populated DarkSide-20k veto tile, including the SiPMs, printed circuit board and front-end electronics, has a measured \(^{40}\mathrm{K}\) activity of \(5.1\pm1.0\,\mathrm{mBq}\)~\cite{veto20k}. This corresponds to \(204\pm40\,\mu\mathrm{Bq/cm^2}\) when normalised to the nominal \(5\times5\,\mathrm{cm^2}\) tile area, or approximately \(230\pm45\,\mu\mathrm{Bq/cm^2}\) when normalised to the $22.18~
\text{cm}^2$ area occupied by its SiPMs. For comparison, the Hamamatsu R11410-21 PMTs characterised for XENON1T~\cite{Aprile_2022} have a reported \(^{40}\mathrm{K}\) activity of \(370\pm60\,\mu\mathrm{Bq/cm^2}\) of effective photocathode area. In terms of U/Th specific radioactivity, similar levels are observed between the veto tile and the most radiopure PMTs used in liquid-xenon experiments. 

Among silicon-based photosensors, SiPMs are particularly attractive because they combine low intrinsic radioactivity with low-voltage operation and, following appropriate technological optimization, excellent performance at cryogenic temperatures. This combination has led several noble-liquid experiments to adopt SiPMs as their baseline photosensor technology. These advantages have motivated the use of SiPMs for liquid-xenon (LXe) scintillation-light detection in the MEG~II experiment, which is currently collecting physics data~\cite{Baldini2018}, as well as their selection as the baseline photosensor technology for the nEXO experiment~\cite{Gallina2022}. SiPMs are also employed in liquid argon (LAr) detector systems for the DUNE experiment~\cite{FALCONE2021164648} and the DarkSide-20k (DS-20k) experiment. DS-20k is a dark matter detector under construction at the Istituto Nazionale di Fisica Nucleare (INFN) Laboratori Nazionali del Gran Sasso (LNGS), designed to enhance the sensitivity for detecting Weakly Interacting Massive Particles (WIMPs). The detector will operate at a level where solar and atmospheric neutrinos become significant backgrounds. The core of DS-20k is a dual-phase Time Projection Chamber (TPC) with a vertical electron drift. The TPC is an octagonal prism (\SI{348}{cm} tall, \SI{350}{cm} in diameter) containing 50 tonnes of underground argon (UAr), with a 20-tonne fiducial volume~\cite{Agnes2024UAr}. Key features include: (i) Electron extraction in the gas phase via a stainless steel grid with 235~$\mu\text{m}$ diameter wires and a 7~mm thick gas pocket; (ii) Acrylic top and bottom lids coated with PEDOT:PSS (Clevios\textsuperscript{\texttrademark}) and tetraphenyl butadiene (TPB) to shift LAr scintillation light from \SI{128}{nm} to visible light and (iii) A field cage made of conductive rings, biased to establish a uniform \SI{200}{V/cm} drift field. Scintillation light is detected by two 10.5~$\text{m}^2$ optical planes equipped with Fondazione Bruno Kessler (FBK) near-ultraviolet high-density (NUV-HD)-cryo SiPMs, positioned at the top and bottom of the TPC to read out the S1 (scintillation) and S2 (electroluminescence) signals. The TPC is enclosed in a stainless steel vessel containing an additional \SI{36}{tonnes} of UAr, which is immersed in \SI{650}{tonnes} of atmospheric liquid argon (AAr) inside a membrane cryostat similar to that adopted for the Deep Underground Neutrino Experiment (DUNE)~\cite{DUNE1}. Both argon volumes are equipped with SiPMs, serving as Inner Veto (IV) and Outer Veto (OV) detectors. The FBK NUV-HD SiPM technology, introduced in 2016~\cite{piemonte2016performance}, faced issues when operated at cryogenic temperatures, {\it e.g.} i) limited Dark Count Rate (DCR) reduction with temperature, due to the onset of a dominant tunneling generation mechanism, and ii) a significant increase in afterpulsing probability, making it unsuitable for a dark matter search~\cite{Aalseth2018}. To address these challenges, FBK developed a low-field (LF) version, referred to as FBK NUV-HD-LF, which reduced the DCR by nearly three orders of magnitude at 77~K~\cite{Acerbi2017}. Further enhancements led to the FBK NUV-HD-cryo technology, the one adopted in DS-20k, which  also improved the afterpulsing probability and microcell recharge time constant at cryogenic temperatures~\cite{Gola2019}. 

\begin{figure}[ht]
\centering
\centering\includegraphics[width=0.99\linewidth]{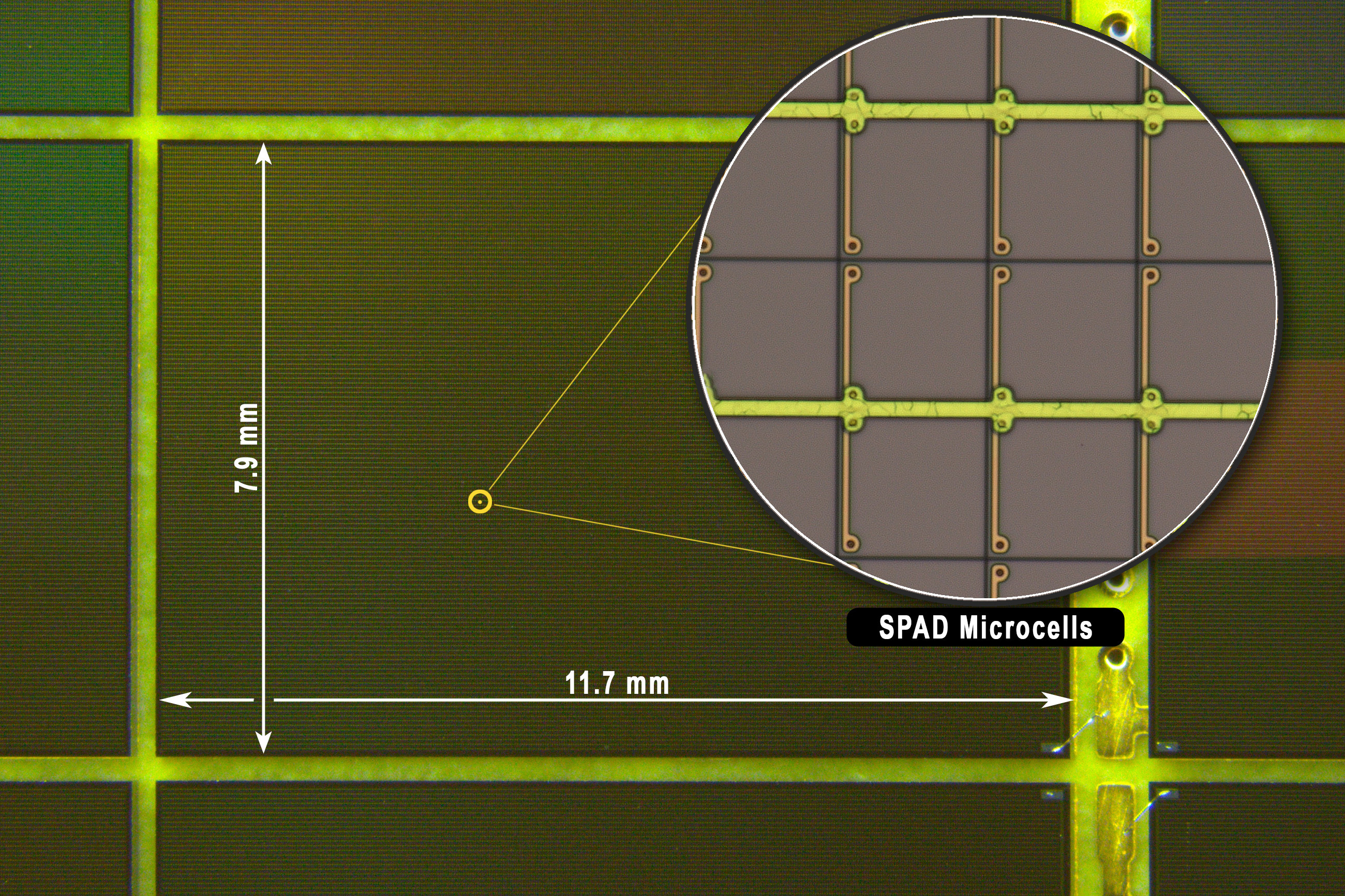}
\caption{FBK NUV-HD-cryo SiPMs. The devices tested have a micro-cell pitch of $30\text{ }\upmu\text{m}$ and a total area of $11.7\times\SI{7.9}{m\meter^2}$.}
\label{fig:SiPM}
\end{figure}

In the final DS-20k design, 24 FBK NUV-HD-cryo SiPMs are grouped into a tile, with each tile containing six branches of four (or two) SiPMs in series, connected to a transimpedance amplifier (TIA) or a custom ASIC amplifier \cite{DIncecco2018,Kugathasan:2020xry}. Four tiles form a quadrant, and four quadrants form a photodetector unit (PDU), with a total surface area of $400~\mathrm{cm^2}$. A total of 680 PDUs are used to instrument the TPC optical planes, IV, and OV detectors, making DS-20k the largest SiPM-based cryogenic detector with over 260,000 SiPMs. The signals from these photodetectors are acquired by a dedicated data acquisition system, described in Ref.~\cite{Acerbi_2026}. In Ref.~\cite{20kcollaboration2024qaqcsipm} we presented  the Quality Assurance and Quality Control (QA/QC) of the FBK NUV-HD-cryo SiPM production wafers manufactured by LFoundry s.r.l. (Avezzano, AQ, Italy), an Italian semiconductor company with the infrastructure to produce a large number of devices for DS-20k \cite{ORGANTINI2020164410}. This paper instead focuses on the characterization of the FBK NUV-HD-cryo SiPM performance at the single-device level. The devices tested, shown in Fig.~\ref{fig:SiPM}, have a micro-cell pitch of $30\text{ }\upmu\text{m}$ and a total area of $11.7\times\SI{7.9}{m\meter^2}$. The SiPM fill factor (FF), defined as the fraction of the total device area occupied by the photosensitive regions, is $76.6\%$.


The gain, dark-count rate and correlated-avalanche properties were studied at \SI{77}{\kelvin} as functions of the applied over voltage using dedicated test setups. The over voltage is defined as $V_{\mathrm{OV}} = V - V_{\mathrm{bd}}$, where $V$ is the reverse-bias voltage and $V_{\mathrm{bd}}$ is the
breakdown voltage. The photon detection efficiency (PDE) was
measured under vacuum at \SI{77}{\kelvin} over the wavelength
range \SIrange{350}{820}{\nano\metre}. Secondary photon emission, which can give rise to external optical crosstalk, was measured under vacuum at \SI{86}{\kelvin} over the wavelength range \SIrange{550}{1020}{\nano\metre}. Overall, the FBK NUV-HD-cryo SiPMs demonstrate exceptional performance as photon detectors for the DS-20k experiment, combining the single photon sensitivity of traditional vacuum photomultiplier tubes with minimal residual natural radioactivity.

\section{Hardware Setups}
\label{sec:DSsetup}

Multiple cryogenic test setups were developed within the DS-20k collaboration to evaluate the performance of individual SiPMs at 77~K. The measurements performed for this study were conducted at TRIUMF, using dry cryostats under vacuum conditions with wideband RF amplifiers~\cite{Gallina2022,gallina2021development}, and at INFN Pisa, where the SiPMs were immersed in liquid nitrogen and read out using a TIA~\cite{RazetoElectronics2018}. These results are compared with previously published measurements performed at INFN LNGS~\cite{Boulay2021,Razeto2023}, and with published results for devices belonging to the same technological generation~\cite{Acerbi2022}. Additional comparisons were made with measurements performed by FBK during preliminary engineering runs of the latest-generation technology. These FBK prototypes consisted of arrays of 64 $1\times1~\mathrm{mm}^2$ SiPMs and were therefore considerably smaller than the $11.7\times7.9~\mathrm{mm}^2$ devices adopted for DS-20k. Moreover, the prototypes did not feature the final Anti-Reflective Coating (ARC), as discussed in Sec.~\ref{sec:PDE}. Their noise properties are nevertheless expected to be representative of the final devices, whereas differences in photon detection efficiency may arise from the different ARC configuration. 

The data acquisition (DAQ) systems used for the measurements performed in this study included a CAEN DT5730B digitizer at TRIUMF and a Teledyne HDO6104A oscilloscope at INFN Pisa. SiPM photon emission was measured separately at 86~K at TRIUMF using a dedicated dry cryostat system, paired with an inverted microscope, a spectrograph, and a cryogenic camera. More details are provided in Sec.~\ref{sec:photonemission}.

\subsection{Collected Data and Trigger Configurations}
\label{subsec:pulsecounting}
The SiPM data were collected according to the scheme described in Ref.~\cite{Gallina2019} and can be categorized into three sub-groups: (1) SiPM dark data, collected at 77~K and at multiple over voltages with the SiPMs shielded from light exposure; (2) SiPM light/lamp data, where the SiPMs were illuminated by an external continuous light source; and (3) SiPM photon emission data, where the SiPMs were exposed to an external laser source, and the emitted light ({\it i.e.} secondary photons) was captured by an infrared camera.

Dark-noise (DN) pulses arise when thermally generated or tunneling-generated electron-hole pairs~\cite{Hurkx1992} trigger an avalanche in a SPAD. The charge produced by a single avalanche in a fully recharged microcell defines the single-Photo-electron Equivalent (1~PE) charge. These pulses are used to determine the single-PE gain and extrapolate the breakdown voltage in Sec.~\ref{sec:gain}, and to characterize uncorrelated and correlated noise in Sec.~\ref{sec:CDP}. For each setup, the DAQ system triggered on individual dark pulses with a DAQ threshold above the noise (Sec.~\ref{sec:gain}, Sec.~\ref{sec:CDP}).  Lamp-driven measurements, {\it i.e.}, PDE data, were collected as a function of the applied over voltage by TRIUMF and FBK at temperatures of 77~K and 300~K, respectively. Overall, the procedures involve the measurement of the photosensors in pulse counting mode, as well as the measurement of their photo-current under illumination. More details are reported in Sec.~\ref{sec:PDE}. SiPM photon emission data were instead measured at TRIUMF by illuminating the SiPM with a 405~nm PicoQuant laser and collecting the subsequent emitted light with a Pylon Camera. Measurements of the SiPM emission are composed of sub-exposures, known as frames, each with an exposure time of 500~s, averaged together. The number of photons detected by the camera is then converted into the number of photons emitted by the SiPMs by correcting for the inefficiency of the apparatus, which includes its optical chain, microscope objectives, mirrors, filters, grating, and camera efficiency.  More details are reported in Sec.~\ref{sec:photonemission}.

\section{Experimental Results}
\label{S:ExpRes}

\subsection{Signal Pulse Analysis Procedure} 
\label{S:fitsection}

\begin{figure}[ht]
\centering\includegraphics[width=0.99\linewidth]{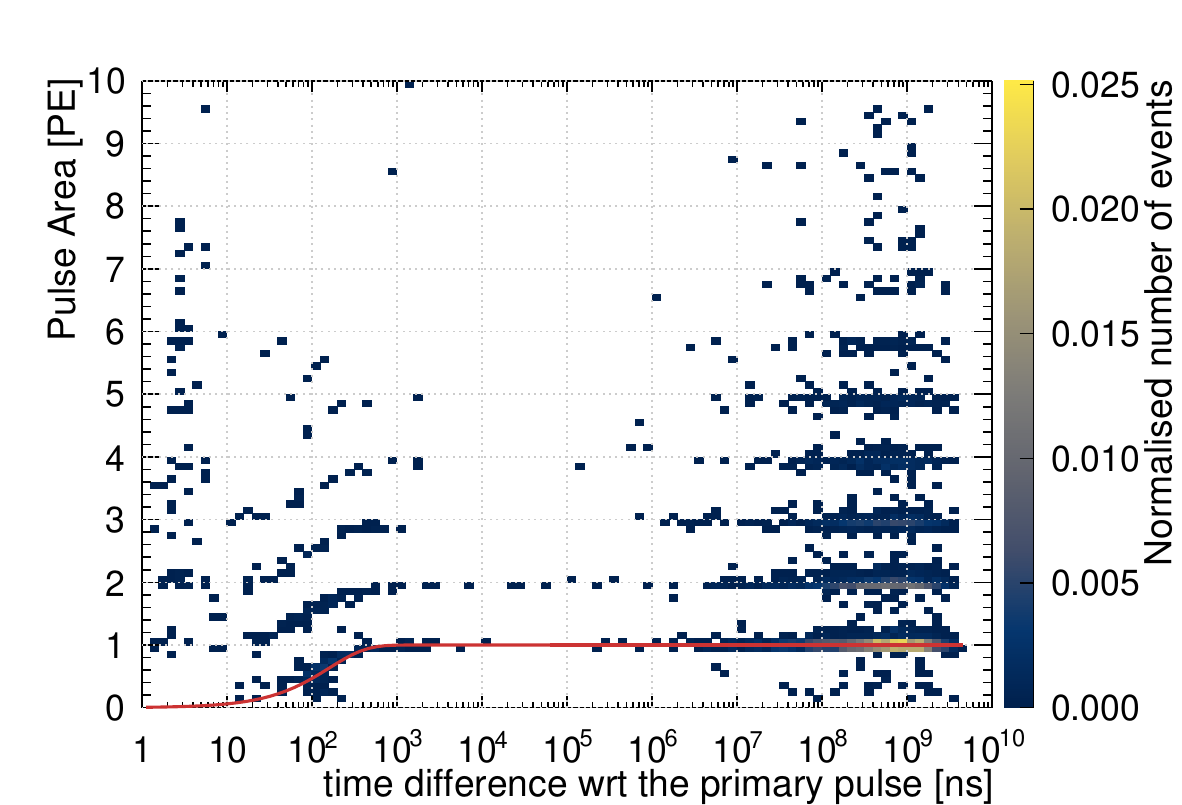}
\caption{Distribution of the pulse area as a function of the time difference, $\Delta t$, between the first pulse following a primary dark-noise trigger and the primary pulse. The measurement was performed at $77$~K and at an over voltage of $10.4\pm0.2$~V. The colour scale represents the number of events in each bin, normalized to the total number of entries. The red curve describes the recovery of a microcell following an avalanche and is parametrized as $A(\Delta t)=A_{\infty}[1-\exp(-\Delta t/\tau_{\mathrm{rec}})]$, with $A_{\infty}=1$~PE. The curve was adjusted to the lower boundary of the single-PE population, yielding an effective recovery time constant of $\tau_{\mathrm{rec}}\simeq160\pm10$~ns. Pulses occurring at short time differences have areas below $1$~PE because the microcell has not yet fully recharged.}
\label{fig:2D}
\end{figure}

The dark data were analyzed at the pulse level by each institution. At TRIUMF, single-PE pulses were fitted using waveform-analysis tools similar to those described in Refs.~\cite{Ako,Gallina2019}, allowing overlapping pulses to be deconvolved and their arrival times and charges to be reconstructed. At INFN Pisa, the pulse charge was instead obtained by summing the baseline-subtracted waveform samples within a fixed integration window around each identified pulse. For isolated pulses, the two methods provide equivalent charge measurements and can therefore be used consistently to determine the SiPM gain (Sec.~\ref{sec:gain}) and extrapolate the breakdown voltage. They also provide equivalent DCR measurements, since the DCR is obtained by counting pulses above a common threshold expressed in PE units, in a time region where pulse overlap is negligible. The waveform-fitting method is, however, particularly useful for the time-resolved Correlated Delayed Avalanches (CDA) analysis, where closely spaced
or overlapping pulses must be separated. For each recorded waveform, the prompt (or primary) pulse is defined as the SiPM pulse that triggers the acquisition. Subsequent pulses are analyzed relative to its arrival time. As an example, we report in Fig.~\ref{fig:2D} the charge distribution of first dark pulses following single PE primary pulses measured by TRIUMF as a function of the time difference with respect to their primary pulse recorded at \SI{77}{\kelvin} and for an over voltage of \SI{10.4(2)}{\volt}. This representation allows the different contributions to the dark signal to be distinguished through their characteristic charge and time distributions. For an illustration in which the different noise components are explicitly labelled, and a discussion of their interpretation, the reader is referred to Ref.~\cite{Ako}; see also Ref.~\cite{gallina2021development}.

\subsection{Single PE Gain and Breakdown Voltage Extrapolation}
\label{sec:gain}

\begin{figure}[ht]
\centering\includegraphics[width=0.99\linewidth]{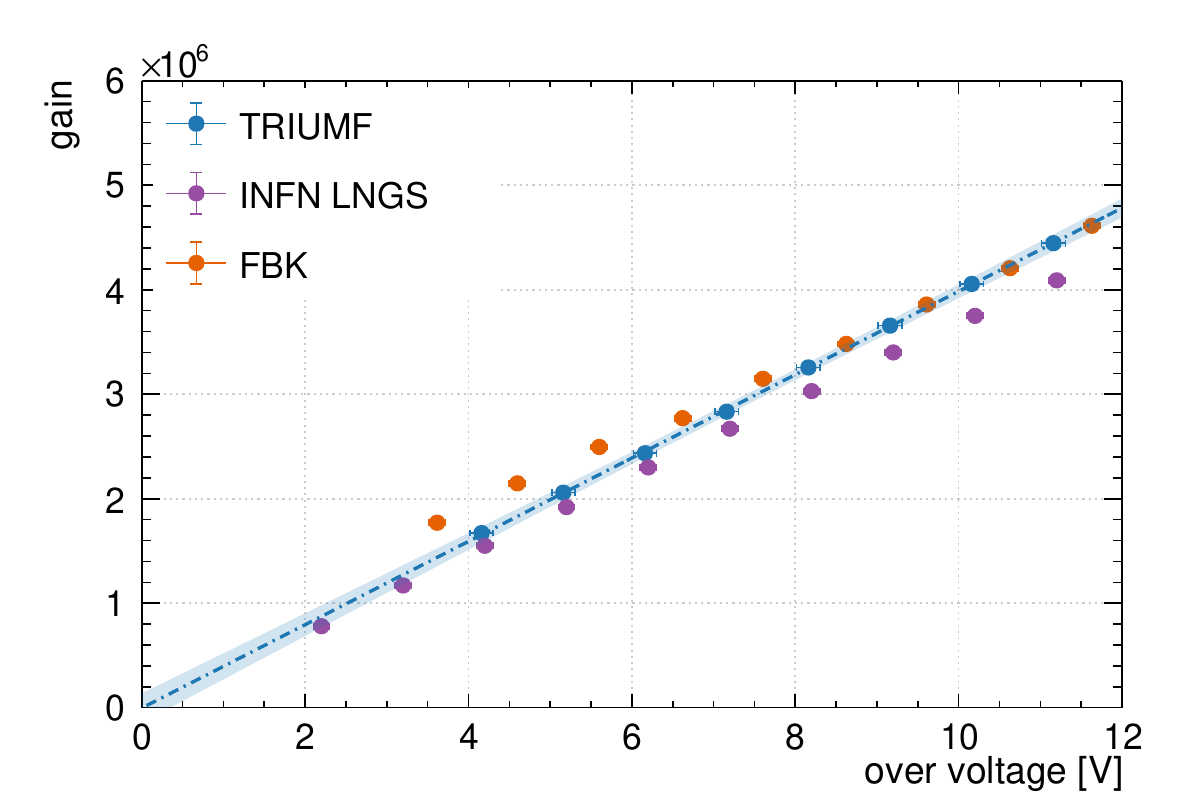}
\caption{Single PE gain as a function of the over voltage measured at 77~K. 
The line represents a linear fit to the TRIUMF data, based on the
expected proportionality between gain and over voltage. The shaded region indicates the corresponding 95\% prediction band. INFN LNGS is the single PE gain previously characterized in Ref.~\cite{Razeto2023}.}
\label{fig:G_OV}
\end{figure}

\begin{figure}[ht]
    \centering
    \includegraphics[width=0.99\linewidth]{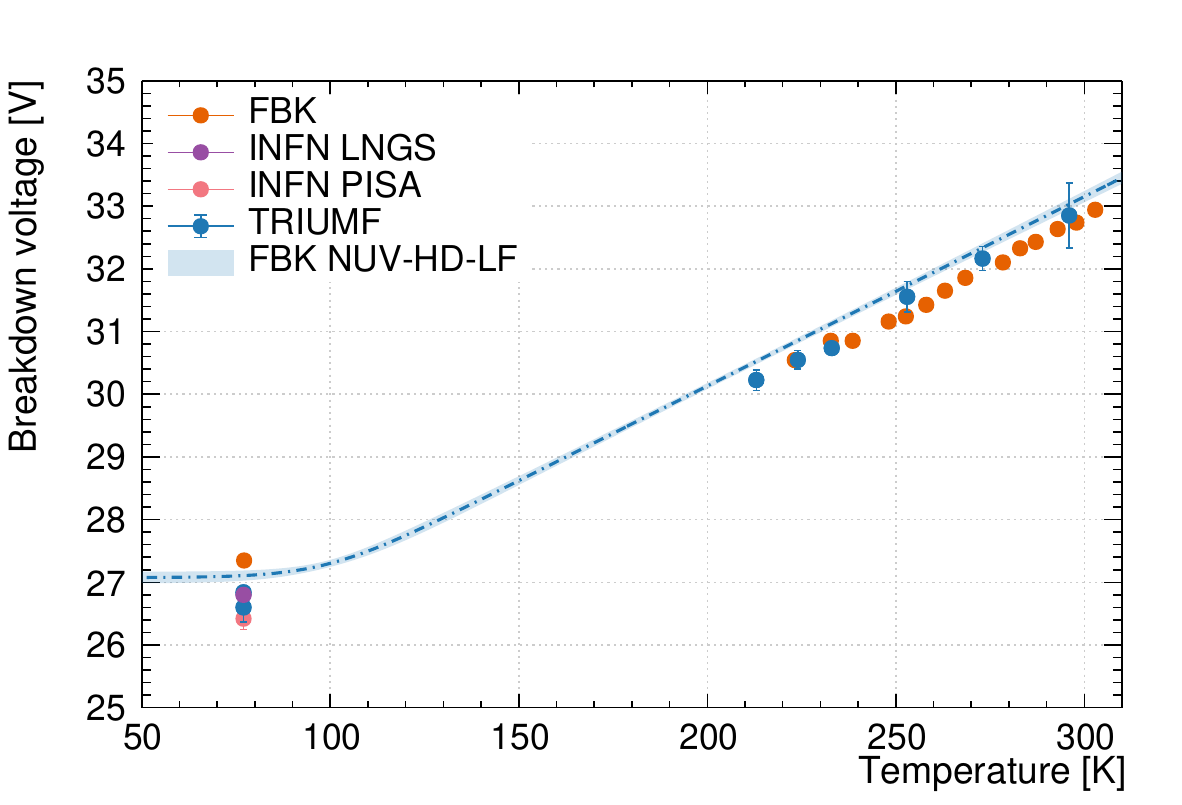}
    \caption{Breakdown voltage as a function of temperature for FBK NUV-HD-cryo SiPMs. The INFN LNGS point corresponds to the value reported in Ref.~\cite{Razeto2023}. The dashed curve is a fit to the previous-generation FBK NUV-HD-LF data reported in Ref.~\cite{Acerbi2017}, while the shaded region represents the corresponding 95\% confidence band.}
    \label{fig:vbd_fbk}
\end{figure}

Single-PE pulses were used to determine the mean single-PE
charge, $Q_{1\mathrm{PE}}$, after correcting the measured
pulse integral for the readout amplification.
The SiPM gain is then given by
\begin{equation}
\label{eq:gain}
G_{1\mathrm{PE}}=\frac{Q_{1\mathrm{PE}}}{q_e},
\end{equation}
where $q_e$ is the elementary charge.
Fig.~\ref{fig:G_OV} shows $\text{G}_{1\text{ PE}}$ as a function of the applied over voltage.
The TRIUMF measurements were fitted with a linear function, shown by the dashed line. 
The linear dependence of the single-PE charge on the applied reverse-bias voltage can be expressed as $Q_{1\mathrm{PE}}=C_D(V-V_{\mathrm{bd}})$, or, equivalently, in terms of the SiPM gain as $G_{1\mathrm{PE}}=(C_D/q_e)(V-V_{\mathrm{bd}})$. The fit therefore provides estimates of (i) the single-SiPM-cell capacitance, $C_D$, from the slope, and (ii) the breakdown voltage, $V_{\mathrm{bd}}$, defined as the bias voltage at which the extrapolated single-PE gain vanishes. The shaded region around the best-fit line represents the 95\% prediction band obtained from the TRIUMF measurements. It includes both the statistical uncertainty of the fitted linear relation and the observed residual dispersion of the individual TRIUMF measurements around the fit. Measurements from FBK and INFN LNGS are shown for comparison. The observed differences among the gain datasets may reflect sample-to-sample variability and systematic differences between the measurement setups. Device-to-device and lot-to-lot variations in breakdown voltage have been quantified for the DS-20k SiPM production in Ref.~\cite{20kcollaboration2024qaqcsipm}, although this study does not directly quantify the gain dispersion at a fixed over voltage. The single-SiPM-cell capacitance, $C_{D}$, determined from the fit, was found to be $63\pm10$~fF. Additionally, the total SiPM capacitance was measured using a capacitance meter (Wayne Kerr 6440B \cite{wayne}), with an extracted single-cell capacitance of $56\pm10$~fF. The breakdown voltage for the device tested by TRIUMF was measured at 77~K to be $26.8\pm0.2$~V, consistent with the results reported in Ref.~\cite{Razeto2023}.  Fig.~\ref{fig:vbd_fbk} shows the SiPM breakdown voltage as a function of temperature. The measurements performed on the devices investigated in this work are compared with those reported in Ref.~\cite{Acerbi2017} for a previous generation of FBK devices (FBK NUV-HD Low-Field technology).

The temperature dependence reported in Ref.~\cite{Acerbi2017} was parametrized using a function that approaches a constant breakdown voltage at low temperature and becomes approximately linear at high temperature. The fit yields a low-temperature plateau of $27.075\pm0.047$~V, a high-temperature coefficient of $30.2\pm0.4$~mV/K, and a transition temperature of $98.9\pm2.7$~K. In the high-temperature region, an independent linear fit to the TRIUMF data gives a temperature coefficient of $32.9\pm3.5$~mV/K, compatible with that obtained for the previous-generation devices. The available TRIUMF measurements at low temperature do not establish the presence of a plateau. Differences of a few tenths of a volt are observed among individual samples and may reflect sample-to-sample variability and differences between production lots. A comparable spread in breakdown voltage among different devices was also reported in Ref.~\cite{20kcollaboration2024qaqcsipm}.

\subsection{Dark Count Rate and Number of Correlated Delayed Avalanches}
\label{sec:CDP}

Unlike the uncorrelated dark-noise pulses introduced in Sec.~\ref{subsec:pulsecounting}, correlated-avalanche (CA) noise is associated with a preceding avalanche and arises primarily from two mechanisms. First, secondary photons emitted during the avalanche in the multiplication region can initiate additional avalanches, giving rise to optically induced CA noise. Second, avalanche-generated charge carriers can become trapped and subsequently be released, triggering delayed avalanches through a process known as afterpulsing. Photons absorbed in the same SiPM can produce internal optical crosstalk or optically induced afterpulsing, while photons escaping from the device can generate external
optical crosstalk in another photosensor. Internal crosstalk avalanches may occur nearly simultaneously with the primary avalanche, referred to as Direct CrossTalk (DiCT), or after a delay of several nanoseconds, referred to as Delayed CrossTalk (DeCT) \cite{Boone2017}. The subset of CAs consisting of afterpulsing and delayed crosstalk events is generally termed Correlated Delayed Avalanches (CDAs). Unlike DN events, CAs (and consequently CDAs) are correlated with a primary signal. DN and CDA are further examined in this section, while secondary photon emission is discussed in more detail in Sec.~\ref{sec:photonemission}. Dark noise pulses can be distinguished from CDA pulses by studying their time distribution relative to the primary pulse, as shown in Ref.~\cite{Butcher2017}. For this analysis, the primary (trigger) pulse was required to have a charge consistent with a single-photoelectron (1 PE) avalanche. The time difference was then calculated between this selected primary pulse and the first subsequent pulse.
\begin{figure}[ht]
\centering
\includegraphics[width=0.99\linewidth]{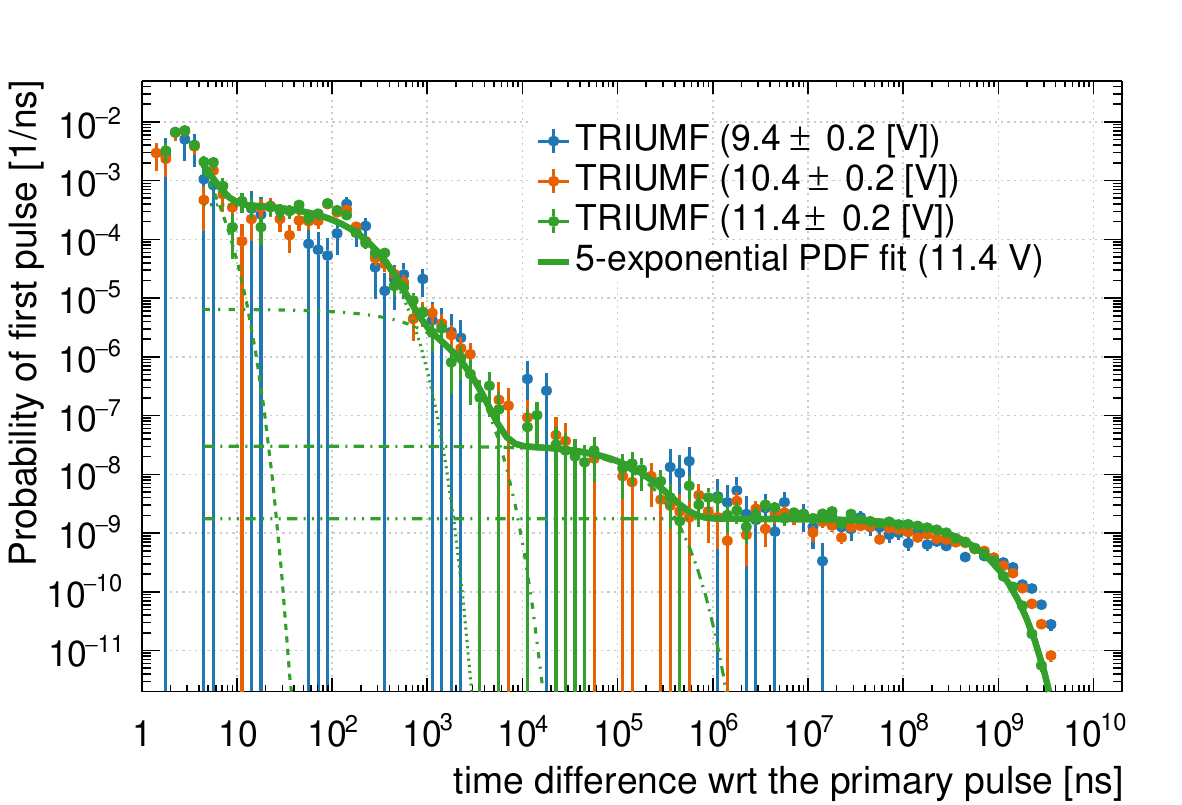}
\includegraphics[width=0.99\linewidth]{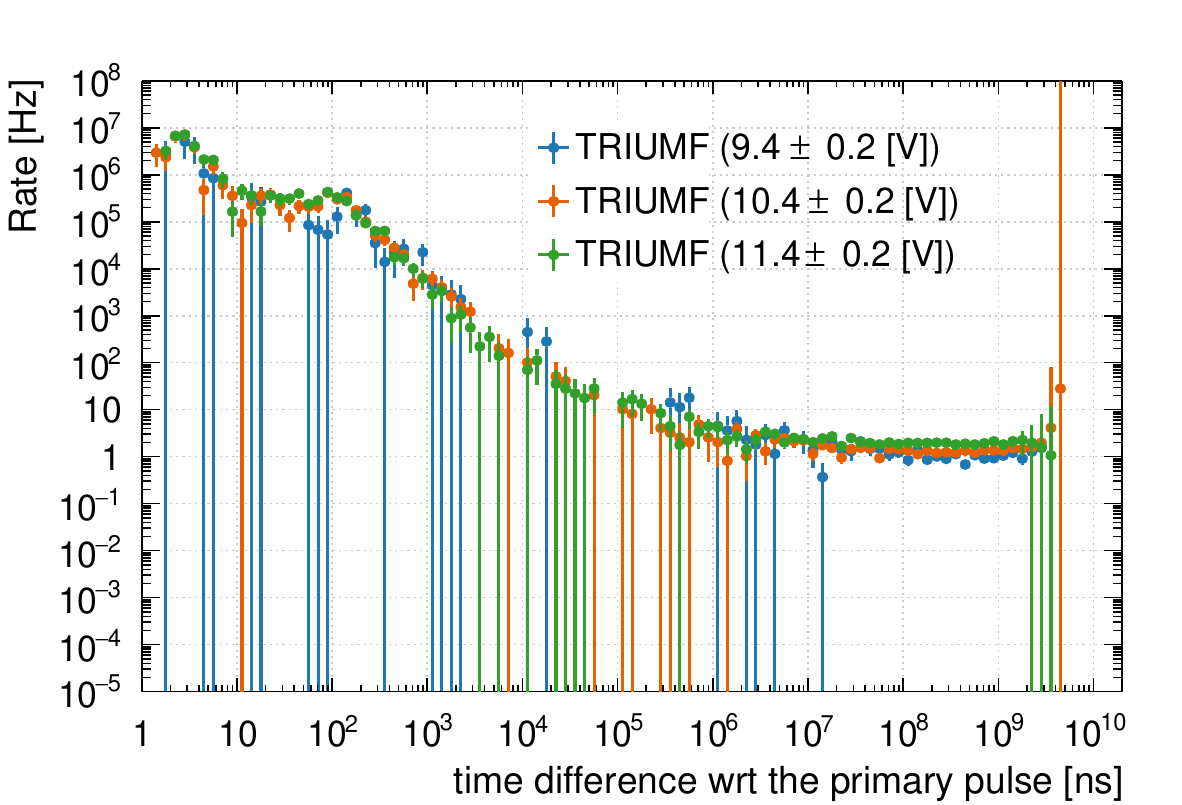}
\caption{(Top) Time distribution of the first secondary pulse following the
primary pulse at 77~K for several over voltages. The solid green line represents the five-exponential fit to the data measured at 11.4~V.
(Bottom) Corresponding secondary-pulse rate $R(t)$. The short-time region is
dominated by correlated delayed avalanches, while the long-time plateau corresponds to the dark-count rate.}
\label{fig:Rate}
\end{figure}

Fig.~\ref{fig:Rate} (top) shows the time distribution of the ﬁrst pulse following the trigger for three over voltages at 77~K. CDA pulses are correlated with the primary avalanche and therefore dominate
the distribution at short and intermediate time differences. Their different characteristic time scales produce the structures and changes of slope observed in the distribution. Conversely, thermally generated dark-noise pulses are uncorrelated with the primary pulse. They consequently follow a Poisson process and give rise to an exponential waiting-time distribution at long time differences. The distribution measured at 11.4~V was fitted with a normalized mixture of
five exponential probability densities: 
\begin{equation}
p(t) =
\sum_{i=1}^{5}
\frac{\alpha_i}{\tau_i}
\exp\left(-\frac{t}{\tau_i}\right),
\qquad
\alpha_i \geq 0,
\qquad
\sum_{i=1}^{5}\alpha_i=1 ,
\label{eq:time_dist_fit}
\end{equation}
where $\tau_i$ is the characteristic time constant of the $i$-th component
and $\alpha_i$ is its relative probability. The first bins below 4~ns were excluded from the fit because they contain the
rising edge of the distribution and can be affected by the pulse-finding
efficiency and the minimum pulse-separation requirement. The fitted parameters are reported in Table~\ref{tab:time_dist_fit}. Four
relatively fast components, with characteristic time constants ranging from
approximately 1.6~ns to 144~$\mu$s, describe the CDA contribution. The
slowest component is attributed to uncorrelated dark-noise pulses. An alternative parameterization, comprising three exponential components and a power-law term ($\propto 1/t^{\alpha}$), describes the data with comparable accuracy. The power-law term accounts for the intermediate time range from \SI{1}{\micro\second} to \SI{1}{\milli\second}, with a fitted exponent of $\alpha=0.96\pm0.05$, in approximate agreement with Ref.~\cite{Itzler10102012}.

\begin{table*}[t]
\centering
\caption{Parameters obtained from the five-exponential fit to the
first-pulse time distribution measured at 11.4~V. Here, $A=11.7\times7.9~\mathrm{mm^2}$ is the total SiPM area used to normalize the rates.}
\label{tab:time_dist_fit}

\begin{tabular}{c c c c}
\hline
Component &
$\alpha_i$ &
$\tau_i$ &
$1/(A\tau_i)$ [Hz/mm$^2$] \\
\hline
1
& 0.0380
& $1.59 \pm 0.22$ ns
& $(6.80 \pm 0.94)\times10^{6}$ \\

2
& 0.0617
& $153.4 \pm 11.2$ ns
& $(7.05 \pm 0.51)\times10^{4}$ \\

3
& 0.00706
& $1096 \pm 393$ ns
& $(9.87 \pm 3.54)\times10^{3}$ \\

4
& 0.00432
& $(1.44 \pm 0.52)\times10^{5}$ ns
& $75.2 \pm 27.2$ \\

5 (DCR)
& 0.8889
& $(5.067 \pm 0.069)\times10^{8}$ ns
& $0.02135 \pm 0.00029$ \\
\hline
\end{tabular}
\end{table*}

\begin{figure}[ht!]
\centering\includegraphics[width=0.99\linewidth]{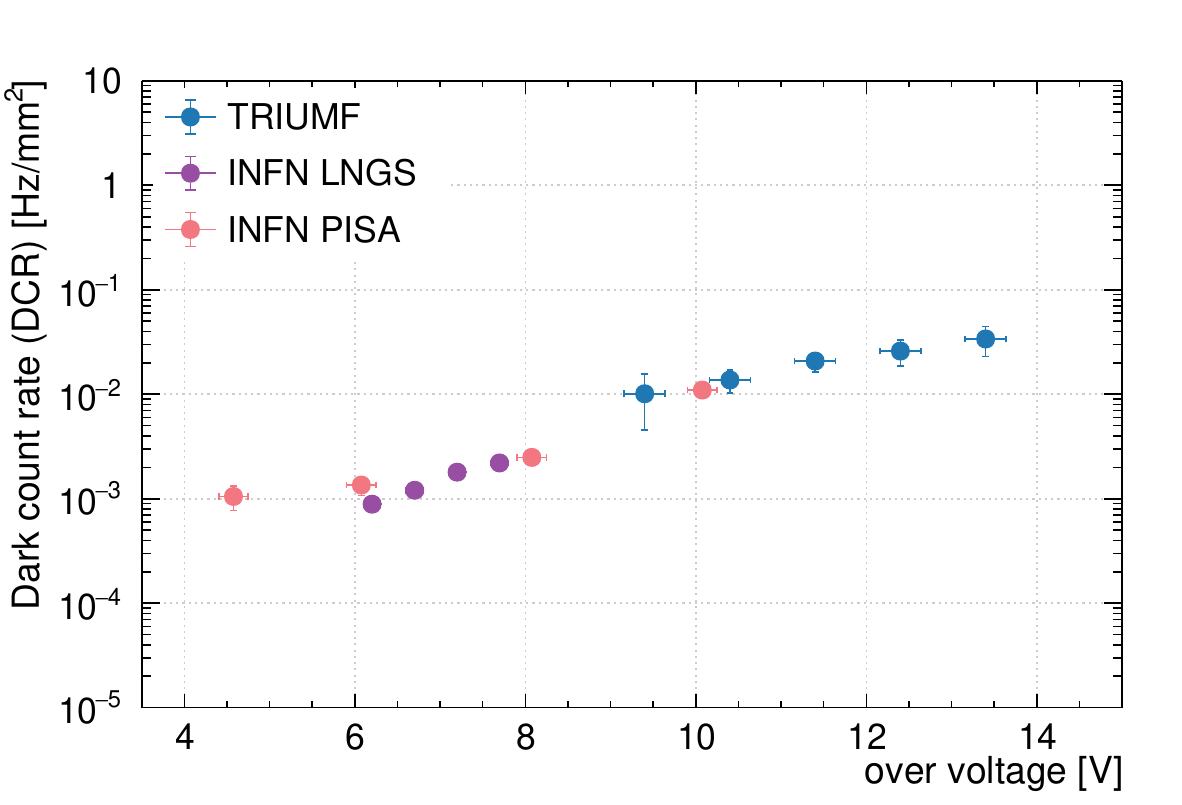}
\caption{Dark Count Rate (DCR) normalized to the total SiPM area, as a function of the applied over voltage, measured at \SI{77}{\kelvin}.}
\label{fig:DN}
\end{figure}

The secondary-pulse rate, $R(t)$, is calculated from the first-pulse time
distribution following the procedure described in
Ref.~\cite{Butcher2017}. It can be expressed as the sum of the dark-count
rate and the rate of correlated delayed avalanches:
\begin{linenomath}
\begin{equation}
R(t)=\text{DCR}+R_{\mathrm{CDA}}(t),
\label{eq:rt}
\end{equation}
\end{linenomath}
where $\text{DCR}$ is the time-independent rate of uncorrelated
dark-noise pulses and $R_{\mathrm{CDA}}(t)$ is the time-dependent rate of
CDAs following the primary pulse. The measured $R(t)$ distributions are
shown in the bottom panel of Fig.~\ref{fig:Rate}. At short time differences, the pulse rate is dominated by CDAs and decreases
rapidly with time. At long time differences, the CDA contribution becomes
negligible and $R(t)$ approaches an approximately constant plateau,
corresponding to the DCR. For each over voltage, the DCR at 77~K is determined from the long-time
plateau of the pulse-rate distribution. Above $10^{7}$~ns, the contribution from correlated delayed avalanches is negligible and the measured rate is approximately constant, as expected for uncorrelated dark-noise pulses. The DCR is calculated as the weighted mean of the rate values in this flat region, with each point weighted by the inverse square of its statistical uncertainty. The DCR uncertainty is taken as the standard error of the
weighted mean. The resulting DCR values, normalized to the total SiPM area, are shown as a function of over voltage in Fig.~\ref{fig:DN}. 

\begin{figure}[ht]
\centering\includegraphics[width=0.99\linewidth]{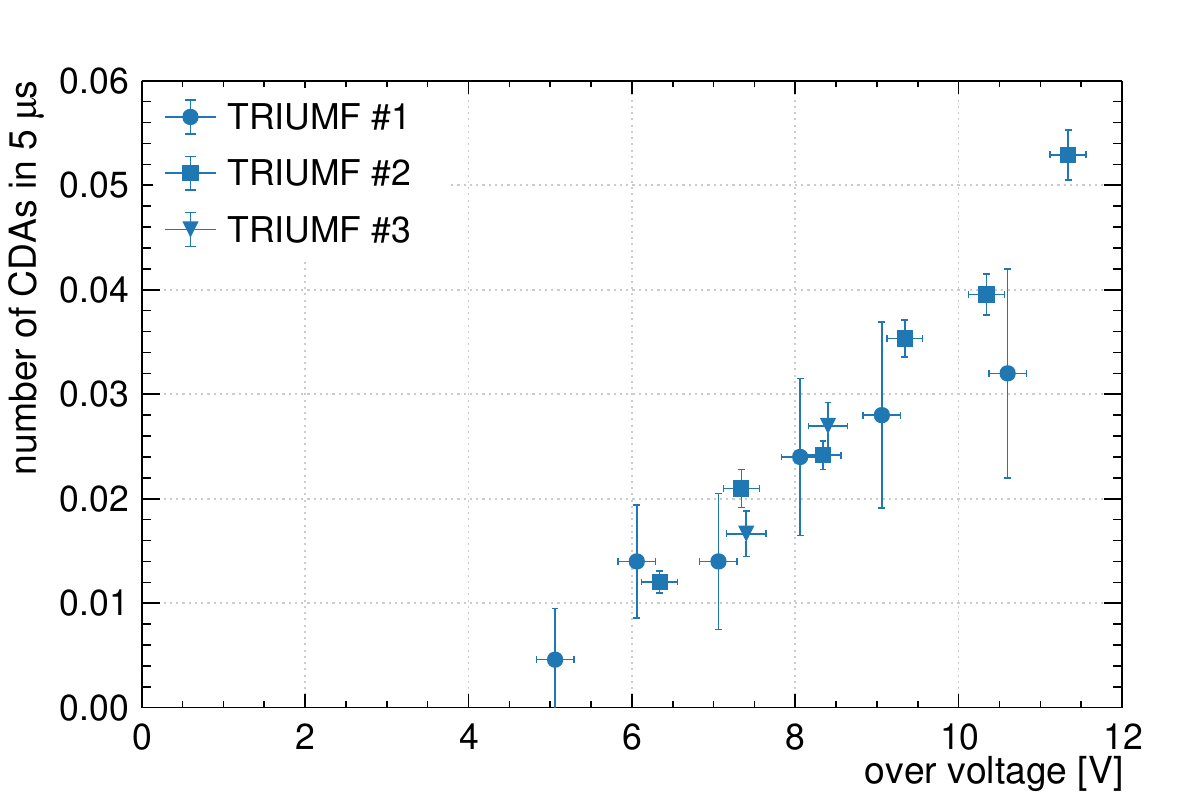}
\caption{Number of Correlated Delayed Avalanches (CDAs) per primary pulse within a time window of $5\mu\text{s}$ after the trigger pulse as a function of the applied over voltage.  Different numbers \#1,\#2,\#3 refer to different SiPMs.}
\label{fig:CDA}
\end{figure}

Fig.~\ref{fig:Rate} (Bottom) can also be used to estimate the number of correlated delayed avalanches per pulse within a fixed time window of length $\Delta t$, here chosen to be $5~\mu\text{s}$. The number of CDAs per pulse can in fact be computed by applying Eq.~\ref{eq:rt} to the observed pulse rate as follows:
\begin{equation}
\label{eq:CDA_eq}
\text{N}_{\text{CDA}}(\Delta t) = \int_{0}^{\Delta t} \Big( \text{R}(t)- \text{DCR} \Big)~\text{dt} 
\end{equation}
where $\text{DCR}$ is reported in Fig.~\ref{fig:DN}. The measured average number of CDAs per pulse in a $5\upmu$s window after the trigger pulse is reported in Fig.~\ref{fig:CDA}, as a function of the applied over voltage.

The $5\upmu$s integration window was selected to match
the time scale relevant for liquid-argon scintillation signals. In particular, it includes both the fast argon component, with a characteristic
time of a few nanoseconds, and more than three time constants of the slow component, whose characteristic time is approximately $1.3$--$1.6\upmu$s. The window therefore contains more than approximately $95\%$ of the slow scintillation component. The measured mean number of CDAs per primary pulse does not vary significantly when the upper limit of the integration window is changed within a reasonable range around $5\upmu$s. Moreover, the average number of CDAs per primary pulse is below approximately $0.06$ over the full investigated over voltage range. This improvement is attributed to FBK's triple doping technology that suppresses SiPM afterpulses, as shown in Ref.~\cite{Capasso2020}. Consequently, this contribution is practically negligible, including any small contribution from CDAs occurring close to or beyond the selected integration boundary.

\subsection{Direct CrossTalk  Analysis}
\label{sec:CT}

\begin{figure}[ht]
\centering
\includegraphics[width=0.99\linewidth]{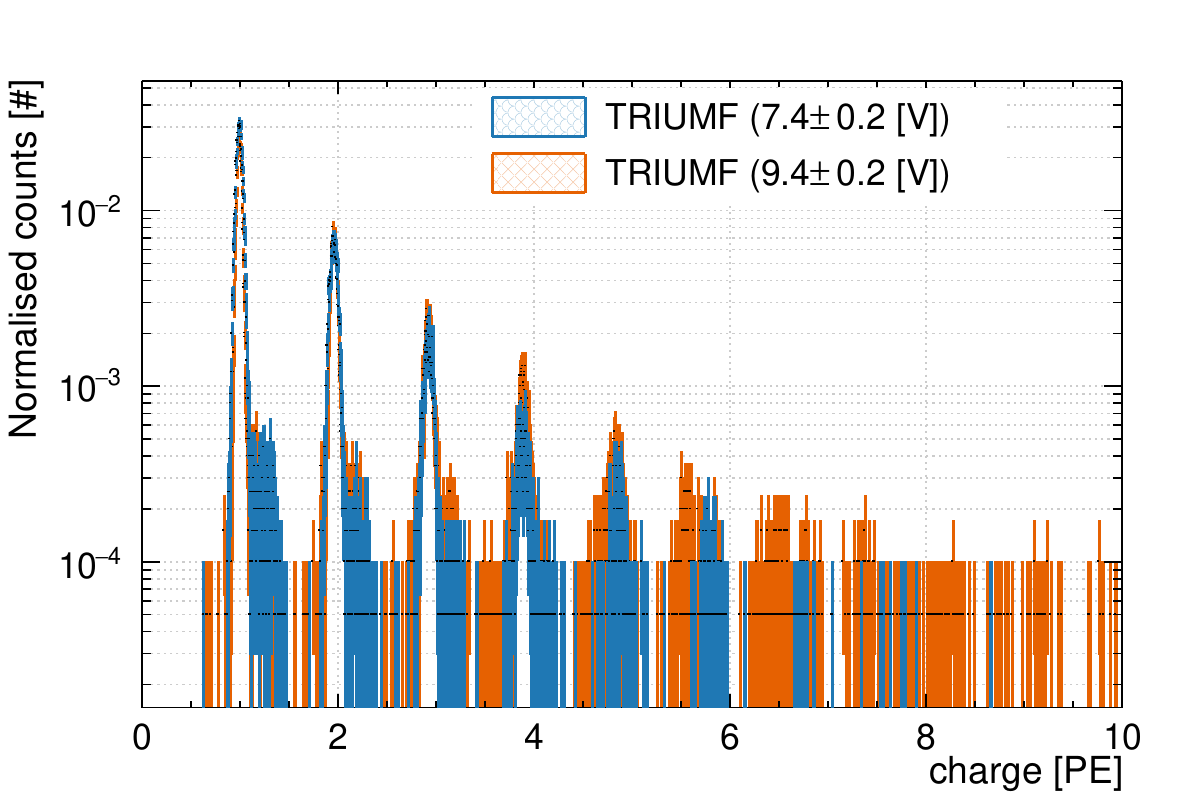}
\caption{Prompt-pulse charge distributions obtained from dark data for FBK
NUV-HD-cryo SiPMs operated at approximately 7 and 9~V of over voltage and at
a temperature of \SI{77}{\kelvin}. The resolved peaks correspond to integer
numbers of simultaneously fired microcells.}
\label{fig:SiPMCharge}
\end{figure}

Based on the dark-count rate reported in Sec.~\ref{sec:CDP}, and assuming Poisson statistics, the probability of two independent dark-noise pulses occurring within the few-nanosecond duration of a prompt pulse is negligible. The charge of the prompt pulse (Sec.~\ref{S:fitsection}) can therefore be used to investigate DiCT, following the approach described in Ref.~\cite{Gallina2019}. Its charge, expressed in PE, provides the total
number of simultaneously fired microcells, including the primary avalanche
and any additional avalanches generated by DiCT. The mean number of
Additional Prompt Avalanches (APAs) is calculated as
\begin{linenomath}
\begin{equation}
N_{\mathrm{APA}}
=
\frac{1}{N}
\sum_{i=1}^{N}
\frac{Q_i}{\overline{Q}_{1\,\mathrm{PE}}}
-1,
\label{eq:napa}
\end{equation}
\end{linenomath}
where $Q_i$ is the charge of the $i$-th prompt pulse,
$\overline{Q}_{1\,\mathrm{PE}}$ is the average single-PE charge defined in
Sec.~\ref{sec:gain}, and $N$ is the total number of analyzed prompt pulses.
The subtraction of one removes the contribution of the primary avalanche. Examples of the prompt-pulse charge distributions measured at approximately
7 and 9~V of over voltage are shown in Fig.~\ref{fig:SiPMCharge}. The
resolved peaks correspond to integer numbers of simultaneously fired
microcells.

\begin{figure}[ht!]
\centering
\includegraphics[width=0.99\linewidth]{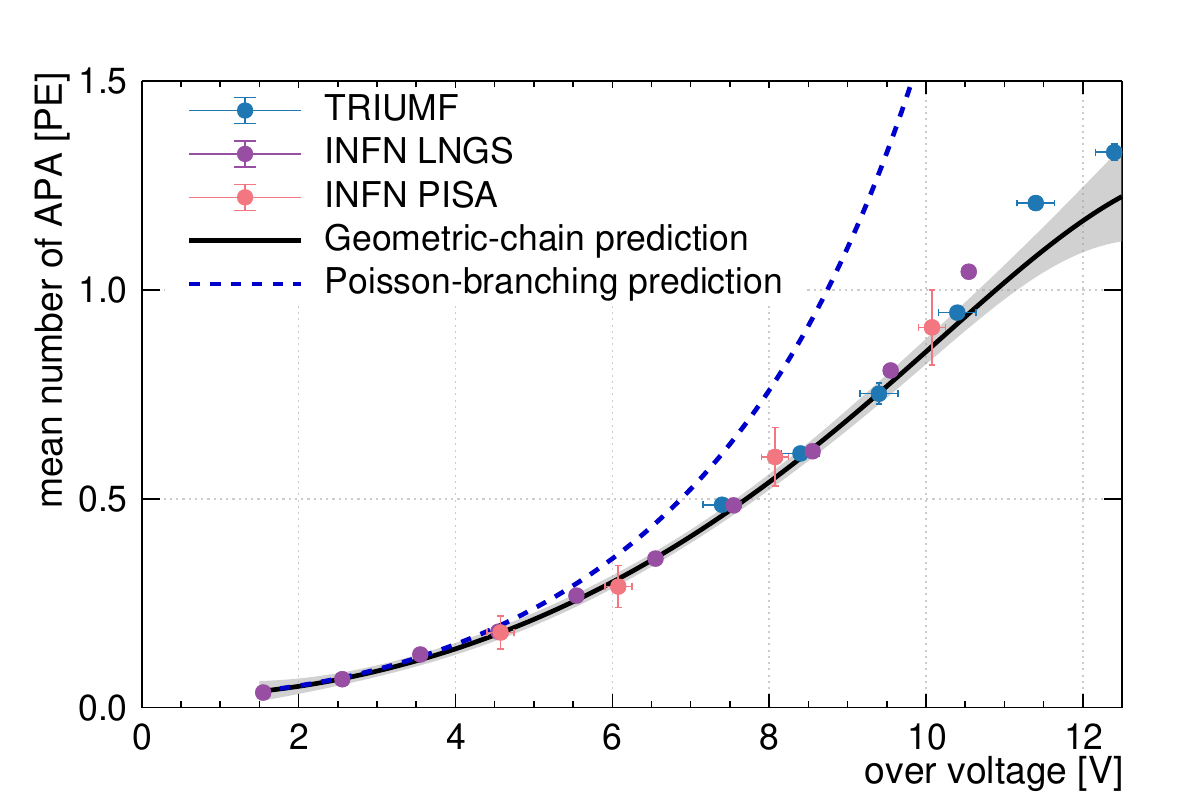}
\caption{Mean number of Additional Prompt Avalanches ($N_{\mathrm{APA}}$)
measured at 77~K as a function of over voltage. The black solid and blue dashed curves represent the geometric-chain and Poisson branching predictions, respectively, both derived from the cubic fit to the DiCT probabilities in
Fig.~\ref{fig:CT_prob}, without fitting the APA data. The shaded region shows the approximate 95\% confidence band propagated from the DiCT fit to the geometric-chain prediction. The INFN LNGS measurements are taken from Ref.~\cite{Boulay2021}.}
\label{fig:APA}
\end{figure}

The resulting $N_{\mathrm{APA}}$, expressed in PE per primary avalanche, is
shown in Fig.~\ref{fig:APA} as a function of the applied over voltage for all
tested devices.

\begin{figure}[ht]
\centering
\includegraphics[width=0.99\linewidth]{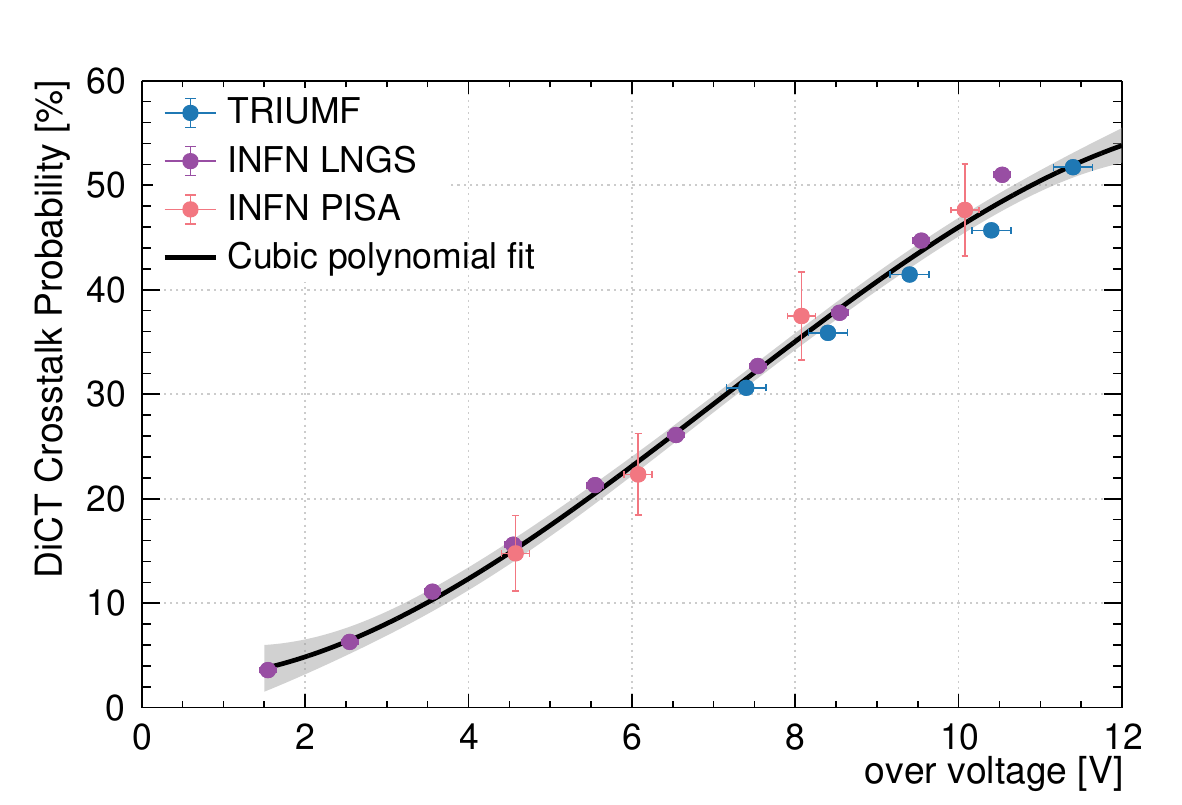}
\caption{Direct crosstalk probability ($P_{\mathrm{CT}}$) measured at 77~K as a function of over voltage. The black curve represents a cubic polynomial fit to the displayed measurements, with equal weight
assigned to each point. The shaded region shows the approximate pointwise 95\% confidence band of the fit. The INFN LNGS measurements are taken from Ref.~\cite{Razeto2023}.}
\label{fig:CT_prob}
\end{figure}

DiCT can be characterized by the probability that a prompt pulse
contains at least one additional avalanche. Experimentally, this probability
is determined as the fraction of prompt pulses with a charge larger than
1.5~PE:
\begin{linenomath}
\begin{equation}
P_{\mathrm{CT}}
=
\frac{N_{Q>1.5\,\mathrm{PE}}}
     {N_{Q>0.5\,\mathrm{PE}}},
\label{eq:pct}
\end{equation}
\end{linenomath}
where the denominator represents the total number of selected prompt pulses.
Unlike $N_{\mathrm{APA}}$, which measures the average additional charge,
$P_{\mathrm{CT}}$ measures only the probability of producing one or more
additional prompt avalanches. The measured DiCT probabilities are shown in Fig.~\ref{fig:CT_prob}. 

The relationship between these two observables can be described using a
geometric-chain model~\cite{VINOGRADOV2012247}. If each avalanche produces a subsequent avalanche with the same probability $p$, the probability of observing a total of $n$ fired microcells is
\begin{linenomath}
\begin{equation}
P(n;p)=(1-p)p^{n-1},
\qquad n\geq1.
\label{eq:geometric_ct}
\end{equation}
\end{linenomath}
In this model, the probability of observing at least one additional
avalanche is
\begin{linenomath}
\begin{equation}
P_{\mathrm{CT}}=P(n\geq2)=p,
\label{eq:pct_p}
\end{equation}
\end{linenomath}
while the mean number of additional avalanches is
\begin{linenomath}
\begin{equation}
N_{\mathrm{APA}}=\langle n\rangle-1=\frac{p}{1-p}=
\frac{P_{\mathrm{CT}}}{1-P_{\mathrm{CT}}}.
\label{eq:napa_pct}
\end{equation}
\end{linenomath}
Equivalently,
\begin{equation}
    P_{\mathrm{CT}}
    = \frac{N_{\mathrm{APA}}}{1+N_{\mathrm{APA}}}.
    \label{eq:geometric_inverse}
\end{equation}
To compare the measurements with this model, the DiCT probability
data were fitted jointly with a cubic polynomial in over voltage $P_{\mathrm{CT}}(V_{\mathrm{OV}})
=a_0+a_1V_{\mathrm{OV}}
+a_2V_{\mathrm{OV}}^2
+a_3V_{\mathrm{OV}}^3$. The fitted probability, shown by the black curve in Fig.~\ref{fig:CT_prob}, was then converted into a prediction for
the mean number of additional prompt avalanches using Eq.~\ref{eq:napa_pct}, as shown by the black curve in Fig.~\ref{fig:APA}. The APA measurements were not
included in the fit and no additional parameters were adjusted
to describe them. The shaded region in Fig.~\ref{fig:CT_prob} represents an approximate pointwise 95\% confidence band. The corresponding band in Fig.~\ref{fig:APA} was obtained by transforming its bounds using Eq.~\ref{eq:napa_pct}. The agreement between the predicted and measured APA values supports the consistency of the two observables with a geometric-chain description over the investigated over voltage range. For comparison, a Poisson branching model was also
considered~\cite{VINOGRADOV2012247}. This prediction is shown by the blue dashed curve in Fig.~\ref{fig:APA}. Although the two predictions approach each other at low crosstalk probability, the Poisson branching model increasingly overestimates
the measured APA values as the over voltage increases.

\subsection{Photon Detection Efficiency}
\label{sec:PDE}

The PDE represents the combined probability that a photon is absorbed within the active volume of the SiPM and subsequently triggers an avalanche. For optimal performance in DS-20k, the selected SiPM must be sensitive to the TPB emission spectrum, which peaks around \SI{420}{\nano \meter} and extends across the [350-600] nm range at LAr temperature~\cite{Francini2013,instruments5010004}. The SiPM spectral response is also relevant to the detection of secondary photons produced during avalanches. These photons can escape the emitting SiPM and trigger avalanches in other photosensors, giving rise to external optical crosstalk, as discussed in Sec.~\ref{sec:CDP}.  SiPM secondary-photon emission extends into the near-infrared region. Over most of the wavelength range investigated in this work (550--1020~nm), the emission intensity increases with wavelength before decreasing near the silicon band-gap limit, as discussed in Sec.~\ref{sec:photonemission}. Photons at longer wavelengths may still be emitted during an avalanche~\cite{mirzoyan2009light}, but they cannot be detected by the SiPM. Moreover, the photon-detection sensitivity progressively decreases as the wavelength approaches this limit. For the devices investigated in this work, the PDE above approximately \SI{820}{\nano\meter} is too small to be measured reliably with the available experimental setup. The PDE characterization was therefore restricted to the wavelength range 350--820~nm. Within this spectral range, the PDE of the FBK SiPMs was measured independently by TRIUMF and FBK. TRIUMF performed the measurements at 77~K in vacuum using the setup described in Ref.~\cite{Gallina2022}, while FBK performed the measurements at 300~K because of limitations of its experimental setup. The temperature dependence of the SiPM PDE is expected to be minimal, as shown in Refs.~\cite{Zhang2022,Acerbi2022}.\\
FBK tested pre-production devices consisting of a matrix of 64 SiPMs, each with an area of $1\times1\,\mathrm{mm}^{2}$, as described in Sec.~\ref{sec:DSsetup}. These devices were coated with a \SI{300}{\micro\metre} thick protective layer with a refractive index of 1.53 and a transmittance greater than 99\% at \SI{400}{\nano\metre}. This differs from the ones tested at TRIUMF that instead featured the final ARC for DS-20k. Despite differences in the tested devices, both institutions used a similar approach: the PDE was calibrated in pulse-counting mode at a single reference wavelength, and its wavelength dependence was then determined from current measurements. This reduces the measurement time compared with repeating the pulse-counting calibration at each wavelength. The PDE was determined at a fixed wavelength of $420$~nm as a function of the applied over voltage by measuring the photon flux detected by the SiPM under illumination ($\text{R}_{\text{0}}$) and dividing it by the absolute photon flux ($\Phi_{0}$) as follows:
\begin{linenomath}
\begin{equation}
\label{eq:PDE_new}
\text{PDE}=\frac{\text{R}_{\text{0}}}{\Phi_{0}}
\end{equation}
\end{linenomath}
with $\Phi_{0}$  defined as 	
\begin{linenomath}
\begin{equation}
\label{eq:flux}
\Phi_{0}=\frac{({I-I_{\text{DCR}}})\lambda}{\text{Rhc}} 
\end{equation}
\end{linenomath}
where $I$ and $I_{\text{DCR}}$ are the photodiode currents with and without illumination, respectively. R is the photodiode responsivity at the wavelength $\lambda$, h is Planck's constant and c is the speed of light. TRIUMF calibrated the light flux using a NIST-calibrated AXUV100G photodiode and a Thorlabs FDS10x10 Si photodiode, depending on the wavelength range, while FBK used calibrated LEDs. No correction for the difference between the photon flux incident on the photodiodes and that incident on the SiPM was required. The two sensors were positioned in the same optical plane, and the light spot was fully contained within the active area of both devices. The light spot was larger than an individual SiPM SPAD and covered several microcells, including the inactive regions between them, so the reported SiPM efficiency already includes the effect of the SiPM fill factor.
\begin{figure}[ht]
\centering\includegraphics[width=0.99\linewidth]{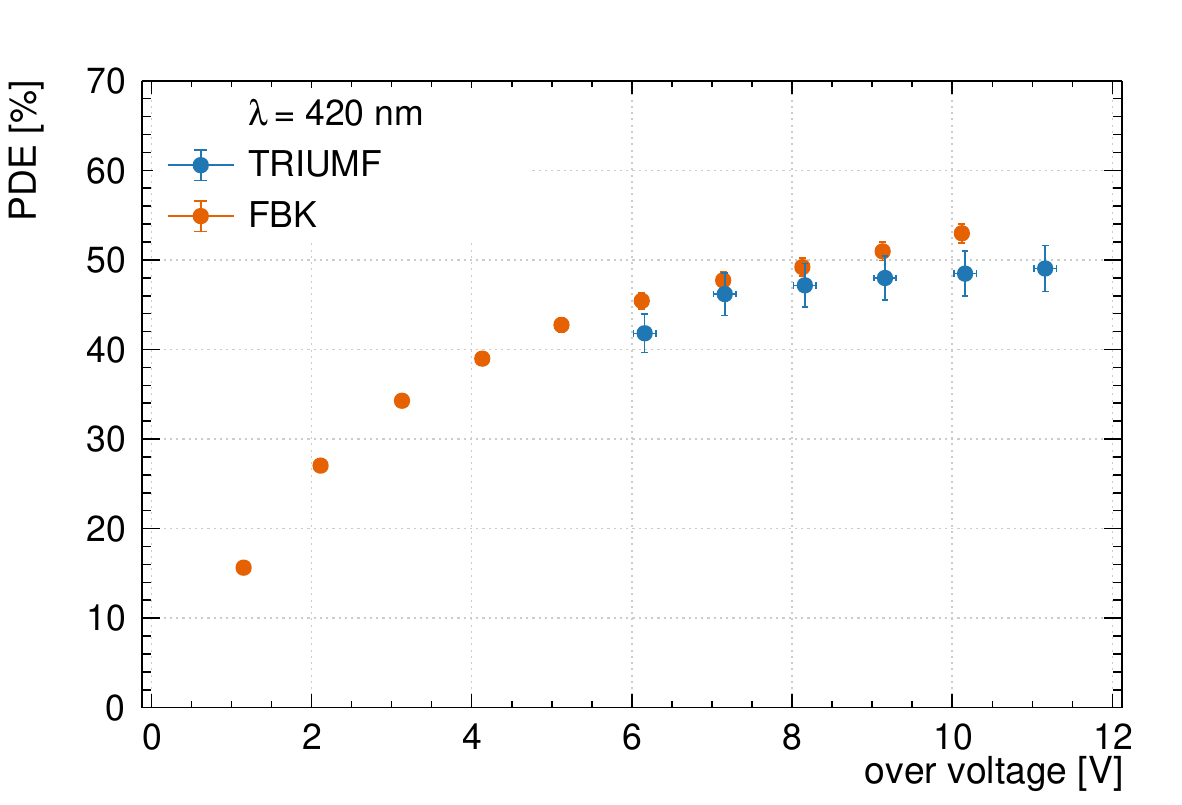}
\caption{Photon Detection Efficiency (PDE) of the FBK NUV-HD-cryo SiPMs measured for a wavelength ($\lambda$) of 420~nm as a function of the applied over voltage. The FBK and TRIUMF measurements were performed at 300~K and 77~K, respectively. The wavelength uncertainty is negligible for the FBK data and approximately 4~nm FWHM for the TRIUMF data. The error bars include both statistical and systematic uncertainties.}
\label{fig:PDE_420_vacuum}
\end{figure}

Fig.~\ref{fig:PDE_420_vacuum} and Table~\ref{tab:PDE_OV} show the PDE measured by TRIUMF at 420~nm as a function of the applied over voltage at 77~K. Fig.~\ref{fig:PDE_420_vacuum} also shows the corresponding measurements performed by FBK at 300~K.  The error bars on each data point represent both statistical and systematic uncertainties. Overall, the two measurements are consistent, within uncertainties.
\begin{table}[ht]
\centering
\caption{Photon Detection Efficiency measured at 77~K and
420~nm as a function of the applied over voltage.
The reported uncertainties are obtained by adding the statistical
and systematic contributions in quadrature.}
\label{tab:PDE_OV}
\small
\begin{tabular}{ccc}
\hline
$V_{\mathrm{OV}}$ [V] &
PDE [\%] &
$\sigma_{\mathrm{tot}}$ [\%] \\
\hline
$6.16 \pm 0.14$  & 41.81 & 2.15 \\
$7.16 \pm 0.14$  & 46.18 & 2.38 \\
$8.16 \pm 0.14$  & 47.16 & 2.43 \\
$9.16 \pm 0.14$  & 47.98 & 2.48 \\
$10.16 \pm 0.14$ & 48.47 & 2.52 \\
$11.16 \pm 0.14$ & 49.04 & 2.55 \\
\hline
\end{tabular}
\end{table}

\begin{figure}[ht]
\centering
\includegraphics[width=0.99\linewidth]{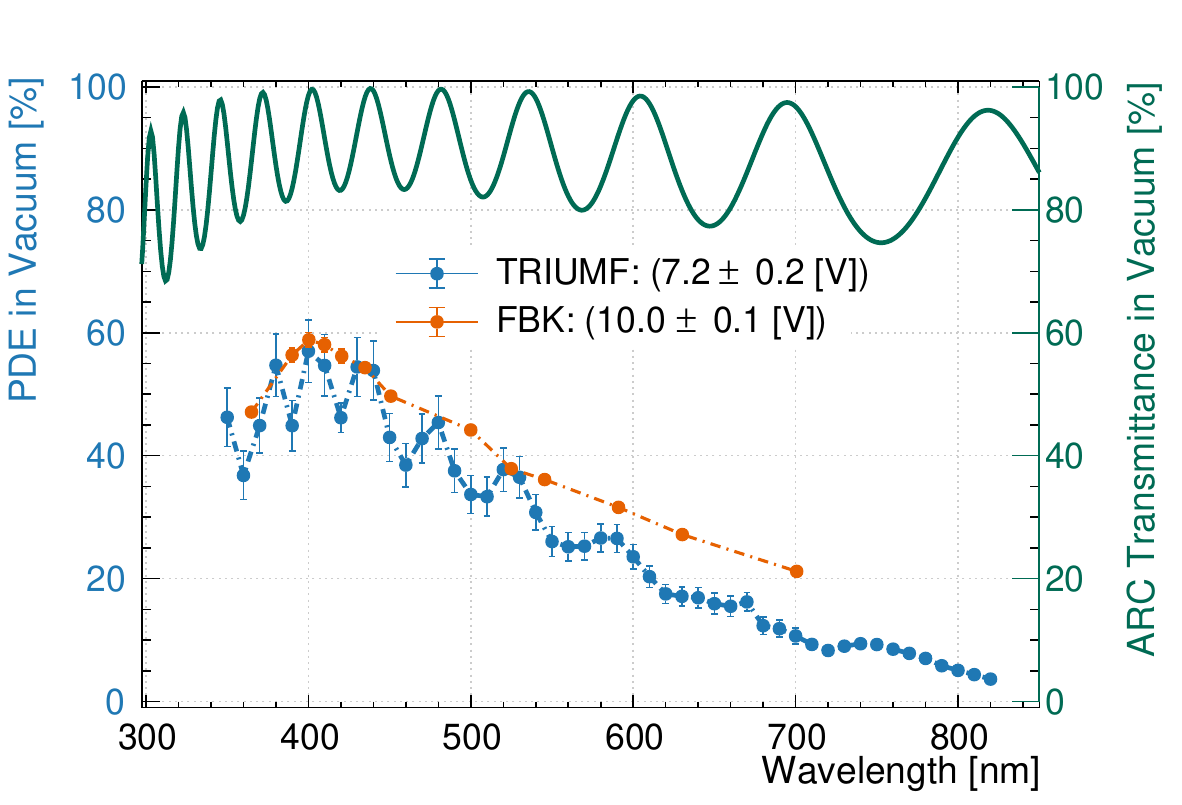}
\caption{Photon Detection Efficiency (PDE) as a function of wavelength. The blue and orange points show measurements performed by TRIUMF at 77~K and by FBK at 300~K. The corresponding over voltages are 7.2~V and 10~V. The green solid curve, referred to the right-hand axis, represents the simulated normal-incidence transmittance of the final antireflective coating (ARC). The wavelength uncertainty is approximately 4~nm FWHM for the TRIUMF data and negligible for the FBK data. The PDE error bars include both statistical and systematic uncertainties.}
\label{fig:PDE_vacuum}
\end{figure}

The wavelength dependence of the SiPM PDE was derived following
the procedure described in Ref.~\cite{Gallina2022} and relying on the correlation between the SiPM current under $420$~nm light illumination and its PDE as follows: 
\begin{linenomath}
\begin{equation}
\label{eq:IHEP_PDE_174}
   \big(I_{\text{SiPM}}(V,420)-I^\text{DCR}_{\text{SiPM}}(V)\big)=\text{PDE}_{420}(V)\times\Phi_0(420)\times f(V)
\end{equation}
\end{linenomath}
where $V$ is the SiPM reverse bias voltage, $I_{\text{SiPM}}$ and $I^\text{DCR}_{\text{SiPM}}$ are the SiPM current with and without illumination, $\text{PDE}_{420}$ and $\Phi_0(420)$ are the SiPM PDE and the photon flux (Eq.~\ref{eq:flux}) measured at $420$~nm with the SiPM and the calibrated diode, respectively.
$f(V)$ is a correction factor that accounts for the SiPM gain and for the CA noise contribution that artificially increases the total current produced by the SiPM (Ref.~\cite{Gallina2022}). Overall $f(V)$ is a function of the applied bias voltage $V$, but can be considered wavelength-independent because it depends only on the SiPM intrinsic characteristics. In this work, the correction factor $f(V)$ is estimated at 77~K using its wavelength and temperature independence by illuminating the SiPM with a pulsed visible light source (420~nm) as follows:
\begin{equation}
f(V)=
\frac{
I_{\mathrm{SiPM}}(V,420)
-
I_{\mathrm{SiPM}}^{\mathrm{DCR}}(V)
}{
\mathrm{PDE}_{420}(V)\,
\Phi_{0}(420)
},
\end{equation}
where $\Phi_{0}$ is the incident photon flux measured with the
calibrated photodiode at 420~nm. The wavelength dependence of the SiPM PDE can be extracted using the PDE measured at $420$~nm and by utilizing a correlation between the SiPM current under illumination and its PDE. More precisely, if the SiPM PDE is known at a specific wavelength ($420$~nm in this case) and for a specific reverse bias voltage $V$, Eq.~\ref{eq:IHEP_PDE_174} can be used to estimate the wavelength-independent correction factor $f(V)$ and to measure the SiPM PDE for the same bias voltage $V$, but for a different wavelength $\lambda$, as follows:
\begin{equation}
\mathrm{PDE}_{\lambda}(V)=
\frac{
I_{\mathrm{SiPM}}(V,\lambda)
-
I_{\mathrm{SiPM}}^{\mathrm{DCR}}(V)
}{
\Phi_{0}(\lambda)\,f(V)
}.
\end{equation}
where $\Phi_{0}(\lambda)$ is defined in Eq.~\ref{eq:flux},  
$I_{\text{SiPM}}$ and $I^\text{DCR}_{\text{SiPM}}$ are the SiPM current with and without illumination. Fig.~\ref{fig:PDE_vacuum} and Table~\ref{tab:PDE_wavelength} show the PDE as a function of wavelength, as measured by TRIUMF at 77~K. Fig.~\ref{fig:PDE_vacuum} also shows the corresponding measurements performed by FBK at 300~K.

Similar trends are observed at the other investigated over voltages. The error bars include both statistical
and systematic uncertainties.
\begin{table}[ht]
\centering
\caption{Selected values of the Photon Detection Efficiency
spectrum measured at 77~K and
$V_{\mathrm{OV}}=7.16\pm0.14$~V.
The reported uncertainties are obtained by adding the statistical
and systematic contributions in quadrature. The wavelength
uncertainty is approximately 4~nm FWHM.}
\label{tab:PDE_wavelength}
\small
\begin{tabular}{ccc}
\hline
$\lambda$ [nm] &
PDE [\%] &
$\sigma_{\mathrm{tot}}$ [\%] \\
\hline
350 & 46.25 & 4.78 \\
400 & 57.01 & 5.08 \\
420 & 46.18 & 2.38 \\
450 & 42.98 & 3.92 \\
500 & 33.69 & 3.11 \\
550 & 26.03 & 2.44 \\
600 & 23.55 & 2.02 \\
650 & 15.93 & 1.72 \\
700 & 10.68 & 1.28 \\
750 &  9.28 & 0.83 \\
800 &  5.07 & 0.51 \\
820 &  3.65 & 0.36 \\
\hline
\end{tabular}
\end{table}
The TRIUMF data exhibit clear oscillations, which are attributed to interference of the incident light within the approximately $1.5~\mu\text{m}$ thick ARC cover layer applied to the surface of these SiPMs. In contrast, the FBK data do not display this interference pattern, as the protective resin applied by FBK to their preproduction samples dampens the oscillations. The resin provides a better refractive index match with the ARC top layer, reducing the interference effect. It is important to note that the \SI{420}{\nano \meter} wavelength corresponds to a minimum in the interference pattern. Compared to the room temperature measurements conducted by FBK, the PDE at 77~K drops by a few percent, particularly in the infrared wavelength range. Similar trends have been observed in other investigations of comparable devices for the same wavelength range~\cite{Zhang2022,Acerbi2022}. Fig.~\ref{fig:PDE_vacuum} also includes a simulation performed by FBK of the ARC transmittance applied to the DS-20k SiPMs, assuming vacuum conditions on the SiPM top surface. The simulation was conducted directly by FBK without disclosing the exact ARC composition. Matching features were also observed in the direct hemispherical reflectance measurement of FBK NUV-HD-cryo SiPMs reported in Ref.~\cite{Boulay2021}. Overall, there is excellent agreement between the TRIUMF data and the simulated transmittance, with both the positions of the maxima and minima of the interference pattern accurately reproduced. However, a deviation is observed in the PDE data at longer wavelengths, likely due to either a degradation in the signal-to-noise ratio of the lamp used by TRIUMF in this wavelength range or a genuine difference in the overall thickness of the ARC stack compared to the one used in the simulation. Notably, the positions of the interference pattern's maxima and minima are highly sensitive to the total thickness of the stack, as discussed in Refs.~\cite{Nakarmi2020,Wagenpfeil2021}.

\subsubsection{Extrapolation of the SiPM PDE in LAr}

In the previous section, we presented the PDE measured at 77~K for FBK NUV-HD-cryo SiPMs under vacuum conditions. Directly measuring the SiPM PDE in LAr, the medium in which the SiPMs will operate in DS-20k, is challenging due to the intrinsic scintillation properties of liquid argon and the non-monochromatic emission spectrum of TPB. In this section, we provide an extrapolation of the expected LAr PDE for these devices by combining simulations performed by FBK, assuming LAr as the medium on the SiPM surface, with other measurements presented in this work. More generally, using the model described in Ref.~\cite{Gallina2019avalanche}, the SiPM PDE can be expressed as
\begin{equation}
\label{eq:PDEvac}
\text{PDE}^{\text{Vac.}}_{\lambda,\theta}(V)=\text{FF}\times\text{T}^{\text{Vac.}}(\lambda,\theta)\times\epsilon_i(V,\lambda)
\end{equation}
Here, FF is the SiPM fill factor, $T^{\mathrm{Vac.}}(\lambda,\theta)$ is the transmittance of the surface coating from vacuum into silicon, and $\epsilon_i(V,\lambda)$ is the effective internal detection probability, accounting for photon absorption in the active volume and subsequent avalanche triggering. The transmittance depends on wavelength, angle of incidence, and the surrounding medium, and includes absorption within the coating. Within this model, $\epsilon_i(V,\lambda)$ is assumed to be independent of the external medium. At normal incidence, $\theta=0$, it can be obtained by
inverting the above expression:
\begin{equation}
\epsilon_i(V,\lambda)= \frac{\text{PDE}^{\text{Vac.}}_{\lambda,0}(V)}{\text{FF}\times\text{T}^{\text{Vac.}}(\lambda,0)}
\end{equation}

\begin{figure}[ht]
\centering
\includegraphics[width=0.99\linewidth]{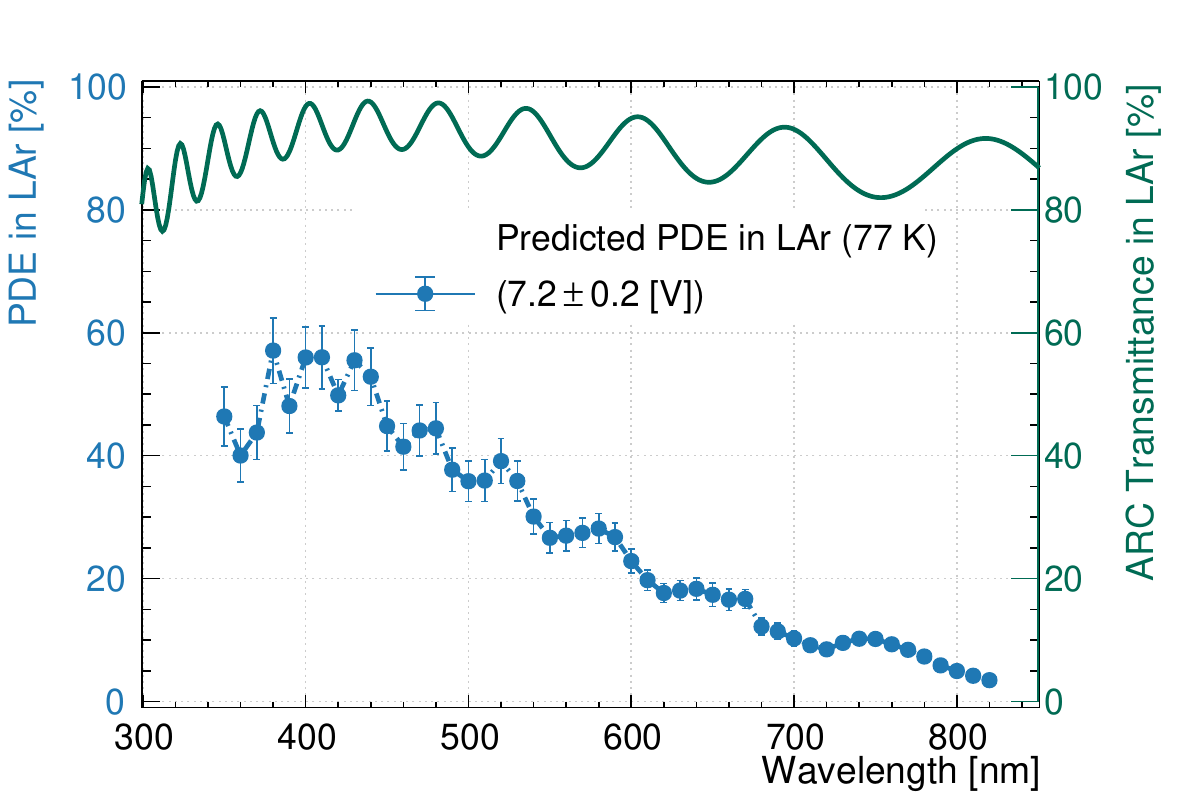}
\caption{Liquid-argon (LAr) Photon Detection Efficiency (PDE) as a function of wavelength. The blue points show the PDE extrapolated from TRIUMF measurements in vacuum at 77~K and an over voltage of 7.2~V, combined with simulations of the SiPM optical response in LAr. The green solid curve, referred to the right-hand axis, represents the simulated normal-incidence antireflective coating (ARC) transmittance in LAr. The PDE error bars include the uncertainties propagated from the vacuum measurements.}
\label{fig:PDE_vacuum_Lar}
\end{figure}

Please note that $\text{PDE}^{\text{Vac.}}_{\lambda,0}(V)$ and $\text{T}^{\text{Vac.}}(\lambda,0)$ are the quantities reported in Fig.~\ref{fig:PDE_vacuum}.  The SiPM PDE in LAr can then be computed by modifying Eq.~\ref{eq:PDEvac}, replacing the simulated transmittance in vacuum with the simulated transmittance in LAr, both at normal incidence, as follows:
\begin{equation}
\begin{aligned} 
\text{PDE}^{\text{LAr}}_{\lambda,0}(V)&=\text{FF}\times\text{T}^{\text{LAr}}(\lambda,0)\times\epsilon_i(V,\lambda)=\\
&=\frac{\text{T}^{\text{LAr}}(\lambda,0)\times\text{PDE}^{\text{Vac.}}_{\lambda,0}(V)}{\text{T}^{\text{Vac.}}(\lambda,0)}
\end{aligned}
\end{equation}
The PDE extrapolated to LAr at normal incidence is shown in Fig.~\ref{fig:PDE_vacuum_Lar}, together with the simulated transmittance. For the transmittance simulation in LAr, we used the refractive index reported in Ref.~\cite{GRACE2017204}. 

\begin{figure}[ht]
\centering
\includegraphics[width=0.99\linewidth]{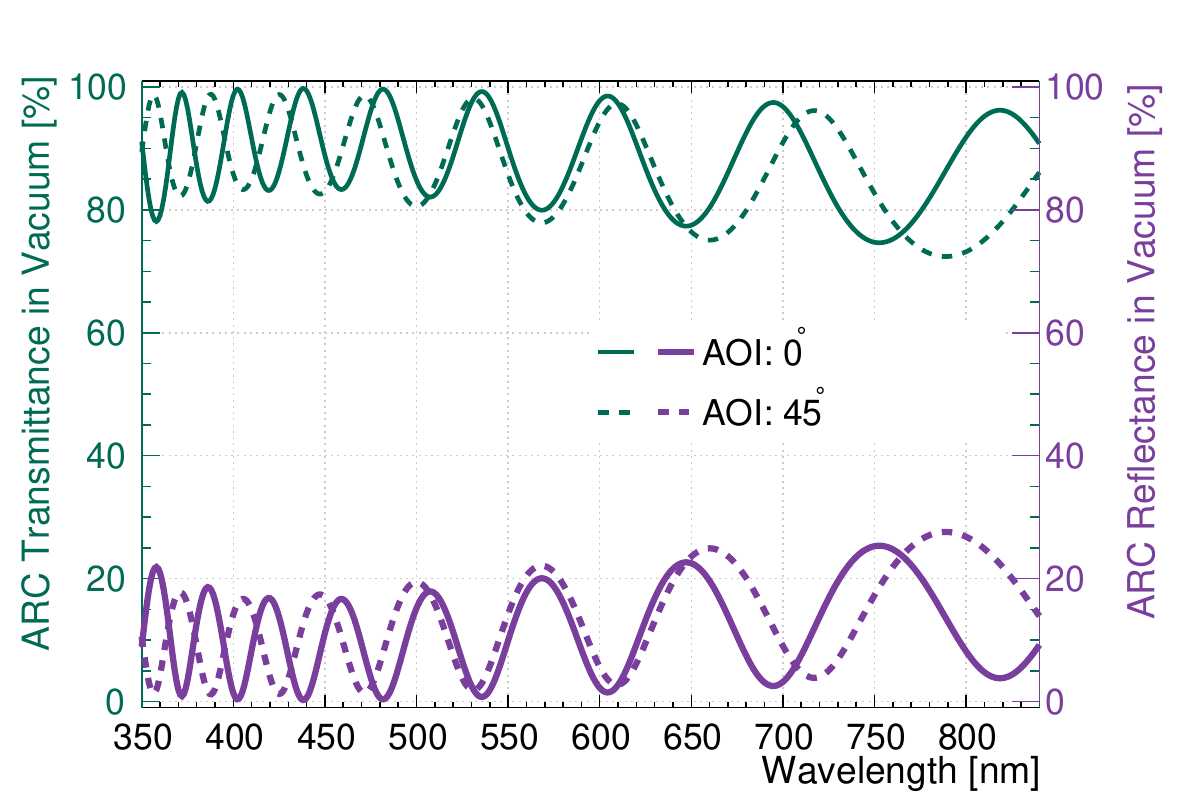}
\includegraphics[width=0.99\linewidth]{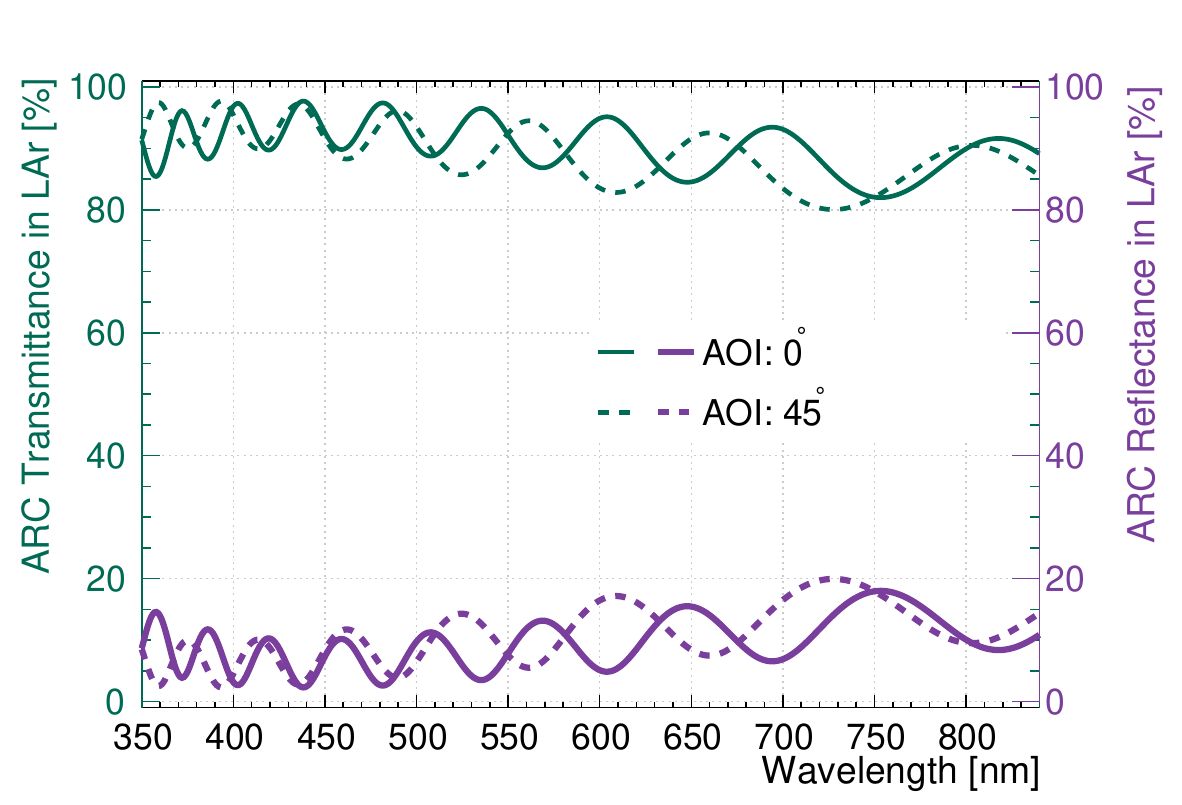}
\caption{Simulated antireflective coating (ARC) transmittance (green curves, left-hand axes) and reflectance (purple curves, right-hand axes) as functions of wavelength in vacuum (top) and liquid argon (bottom). Solid and dashed curves correspond to angles of incidence (AOI) of $0^\circ$ and $45^\circ$, respectively.}
\label{fig:Simulation}
\end{figure}

For completeness, Fig.~\ref{fig:Simulation} shows the simulated transmittance and reflectance in vacuum and LAr as functions of wavelength, for angles of incidence of $0^\circ$ and $45^\circ$. Compared with the simulations in vacuum, the peak-to-valley oscillations in the simulated transmittance, and consequently in the extracted PDE, are significantly reduced. This improvement is due to a better match between the refractive index of LAr and the ARC.

\subsection{SiPM Secondary Photon Emission}
\label{sec:photonemission}

\begin{figure}[ht]
\centering
\includegraphics[width=0.99\linewidth]{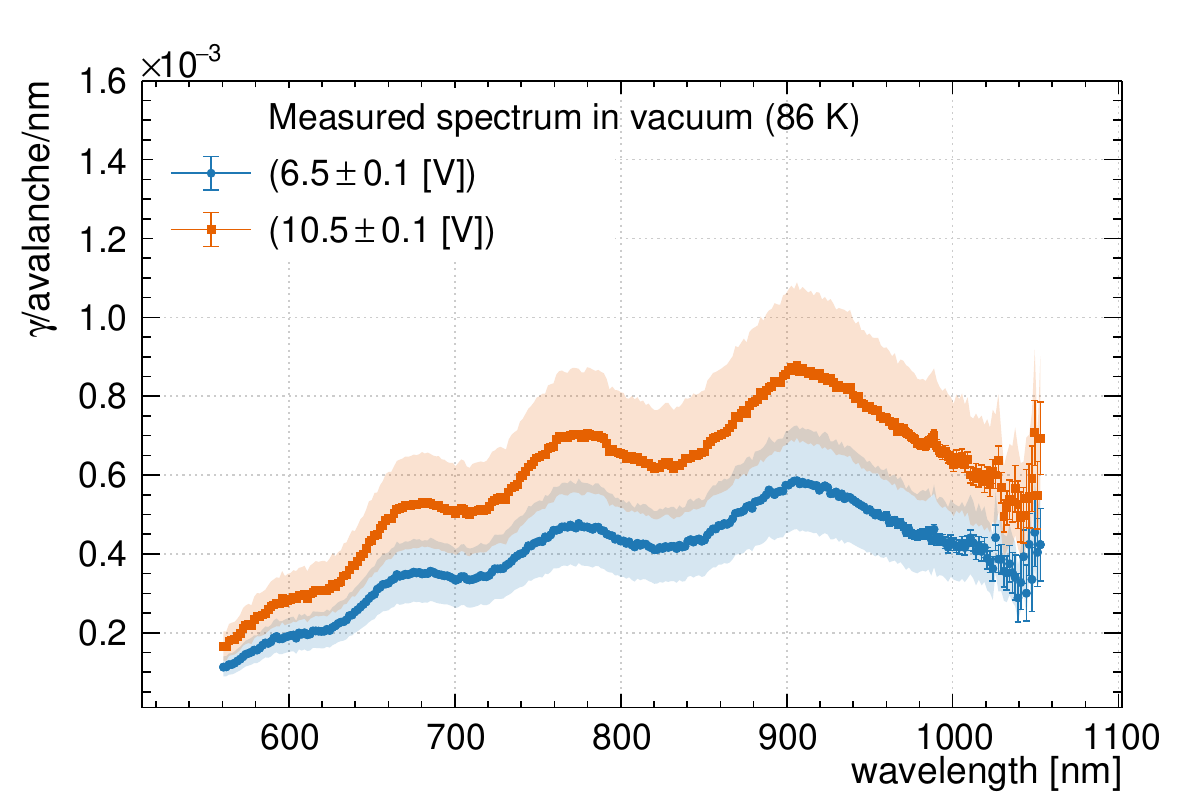}
\caption{Secondary-photon emission spectra $ n_{\gamma}^{\mathrm{Vac.}}(\lambda)$ collected within the acceptance of the microscope objective under vacuum at 86~K, shown for over voltages of 6.5 and 10.5~V. The spectra are expressed as photons per avalanche per nanometre.}
\label{fig:eCT_scan}
\end{figure}

In Sec.~\ref{sec:CDP}, we introduced secondary-photon emission as an intrinsic by-product of the avalanche process. A fraction of these photons can leave the emitting SiPM and trigger avalanches in other photosensors, thereby producing external optical crosstalk. This effect is particularly relevant for large-area SiPM-based detectors such as DS-20k, in which many photosensors are in direct or reflected optical communication. The emission spectrum lies predominantly in the near-infrared and, over most of the wavelength range considered here, its intensity increases with wavelength before decreasing near the silicon band-gap limit. A realistic simulation of external crosstalk therefore requires both the spectral distribution and the absolute photon yield per avalanche. The secondary-emission measurements were performed with the setup described in Ref.~\cite{10713286}. The apparatus is based on an Olympus IX-83 inverted microscope equipped with a vacuum chamber in which the SiPM was cooled to 86~K. The absolute temperature has an estimated systematic uncertainty of up to 10~K, with the SiPM possibly being warmer than the nominal value. However, the temperature stability during data taking was better than 1~K. A pulsed laser was focused onto an individual SPAD through a $20\times$ objective with a numerical aperture (NA) of 0.45. Reflected laser light was rejected by a dichroic mirror and an additional long-pass filter. Light emitted by the SiPM was directed to a Princeton Instruments HRS 300 spectrograph coupled to a Princeton Instruments PyLoN 400BR eXcelon cryogenic camera with a backside-illuminated CCD sensor. In spectroscopy mode, the 400 pixels along the vertical direction retain spatial information about the illuminated SPAD, whereas the 1340 pixels along the horizontal direction sample the calibrated wavelength axis. After background subtraction and image processing, the camera signal was converted into the number of detected photoelectrons in each wavelength bin, $N_{\gamma}^{c}(\lambda)$. The superscript $c$ indicates that only the contribution associated with the central, laser-illuminated SPAD was selected. To obtain the spectrum collected within the acceptance of the microscope objective, this quantity was corrected for the wavelength-dependent efficiency of the complete optical system, $\eta_{\mathrm{sys}}(\lambda)$, and normalized to the number of avalanches occurring in the central SPAD during the exposure, $N_{\mathrm{av}}^{c}$. Expressed per unit wavelength, the result is
\begin{equation}
\label{eq:photon_counting}
    n_{\gamma}^{\mathrm{Vac.}}(\lambda)=
    \frac{N_{\gamma}^{c}(\lambda)}
    {\eta_{\mathrm{sys}}(\lambda)\,N_{\mathrm{av}}^{c}\,\Delta\lambda},
\end{equation}
where $\Delta\lambda$ is the wavelength-bin width, which varies from 1.2 to 2.2~nm because of the nonlinear wavelength calibration. Thus, $n_{\gamma}^{\mathrm{Vac.}}(\lambda)$ is the number of photons per avalanche per nanometre that reach the microscope objective under vacuum conditions; it is not the total number of photons produced in the avalanche or escaping from the SiPM. Figure~\ref{fig:eCT_scan} shows $n_{\gamma}^{\mathrm{Vac.}}(\lambda)$ at over voltages of 6.5 and 10.5~V. The spectra measured at the other over voltages exhibit similar behaviour. The shaded bands represent the systematic uncertainties.

As observed in the PDE measurements shown in Fig.~\ref{fig:PDE_vacuum}, the measured emission spectra contain oscillations caused by optical interference in the micrometre-scale coating layers of the device. Consequently, the spectrum collected in vacuum cannot be converted into the photon yield escaping into liquid argon by applying only a geometric solid-angle correction. The wavelength and angle-dependent transmission of the silicon and of the surface coating must also be included. The conversion is performed in two steps. First, the photon spectrum produced at the avalanche region is inferred from the spectrum measured through the objective in vacuum. Second, the fraction of this source spectrum that is transmitted through the SiPM surface into LAr is calculated. Assuming that the avalanche region is an isotropic point source, these two steps can be combined as
\begin{align}
\label{eq:source_photon_counting}
    n_{\gamma}^{\mathrm{LAr}}(\lambda)
    &=\frac{A^{\mathrm{LAr}}(\lambda)}
         {A^{\mathrm{Vac.}}(\lambda)}
    \,n_{\gamma}^{\mathrm{Vac.}}(\lambda)=\\
    &=\frac{A^{\mathrm{LAr}}(\lambda)}
         {A^{\mathrm{Vac.}}(\lambda)}
    \frac{N_{\gamma}^{c}(\lambda)}
         {\eta_{\mathrm{sys}}(\lambda)\,N_{\mathrm{av}}^{c}\,\Delta\lambda}.
\end{align}
Here, $A^{\mathrm{Vac.}}(\lambda)$ is the fraction of isotropically produced photons that reaches the microscope objective in the vacuum measurement, whereas $A^{\mathrm{LAr}}(\lambda)$ is the fraction that escapes from the SiPM into the LAr half-space. For either configuration,
\begin{align}
\label{eq:SiPM_Transmission_Vac}
    A^{m}(\lambda)
    =
    \frac{
    \displaystyle \int_{0}^{2\pi} d\phi
    \int_{0}^{\theta_{\mathrm{Si}}^{m}(\lambda)}
    T_{\mathrm{SiPM}}^{m}(\lambda,\theta)\,
    \sin\theta\,d\theta}
    {\displaystyle \int_{0}^{2\pi} d\phi
    \int_{0}^{\pi}\sin\theta\,d\theta},
\end{align}
with $m\in\{\mathrm{Vac.},\mathrm{LAr}\}$. The denominator is the full solid angle of the assumed isotropic source. The upper integration limit, $\theta_{\mathrm{Si}}^{m}(\lambda)$, is expressed as an emission angle inside silicon and depends on the final optical acceptance. For the vacuum measurement, the objective NA gives
\begin{align}
    &n_{\mathrm{Si}}(\lambda)\,
    \sin\theta_{\mathrm{Si}}^{\mathrm{Vac.}}(\lambda)
    = \mathrm{NA}\\
    &\theta_{\mathrm{Si}}^{\mathrm{Vac.}}(\lambda)
    =
    \sin^{-1}\!\left[
    \frac{\mathrm{NA}}{n_{\mathrm{Si}}(\lambda)}
    \right]
\end{align}
For emission into LAr, all rays transmitted into the outward LAr hemisphere are included. The limiting ray propagates parallel to the interface in LAr, so Snell's law gives
\begin{align}
    &n_{\mathrm{LAr}}(\lambda)\sin\left(\frac{\pi}{2}\right)
    =
    n_{\mathrm{Si}}(\lambda)
    \sin\theta_{\mathrm{Si}}^{\mathrm{LAr}}(\lambda)\\
    &\theta_{\mathrm{Si}}^{\mathrm{LAr}}(\lambda)
    =
    \sin^{-1}\!\left[
    \frac{n_{\mathrm{LAr}}(\lambda)}
         {n_{\mathrm{Si}}(\lambda)}
    \right].
\end{align}
The transmission probability entering Eq.~\ref{eq:SiPM_Transmission_Vac} is written as
\begin{equation}
\label{eq:SiPM_Transmission}
    T_{\mathrm{SiPM}}^{m}(\lambda,\theta)
    =
    T_{\mathrm{Si}}(\lambda,\theta)\,
    T_{\mathrm{ARC}}^{m}(\lambda,\theta)
\end{equation}
with $ m\in\{\mathrm{Vac.},\mathrm{LAr}\}$. Here, $T_{\mathrm{Si}}(\lambda,\theta)$ accounts for absorption along the path from the avalanche region to the silicon surface, while $T_{\mathrm{ARC}}^{m}(\lambda,\theta)$ describes transmission through the surface-coating stack into the external medium $m$. As in Ref.~\cite{10713286}, the \(A\) factors were calculated using a simplified optical model in which the SiPM surface consists of a single SiO\(_2\) layer deposited on the silicon substrate. This single-layer approximation cannot reproduce the full optical response of the actual multilayer anti-reflective coating (ARC), including the spectral oscillations discussed in Sec.~\ref{sec:PDE}. For example, varying the assumed effective SiO$_2$ thickness by $\pm0.02\mu$m shifts a representative simulated transmission maximum by approximately $\pm8$nm, thereby modifying the wavelength-dependent correction applied to the measured spectrum. A more realistic optical model could not be implemented because the composition and thicknesses of the individual ARC layers are proprietary and were not disclosed by FBK. The resulting $ A^{m}(\lambda)$ factors, and hence the photon emission extrapolated to LAr, are therefore subject to an additional, unquantified systematic uncertainty associated with the assumed coating description, which is not included in the reported uncertainty bands.

\begin{figure}[ht]
\centering
\includegraphics[width=0.99\linewidth]{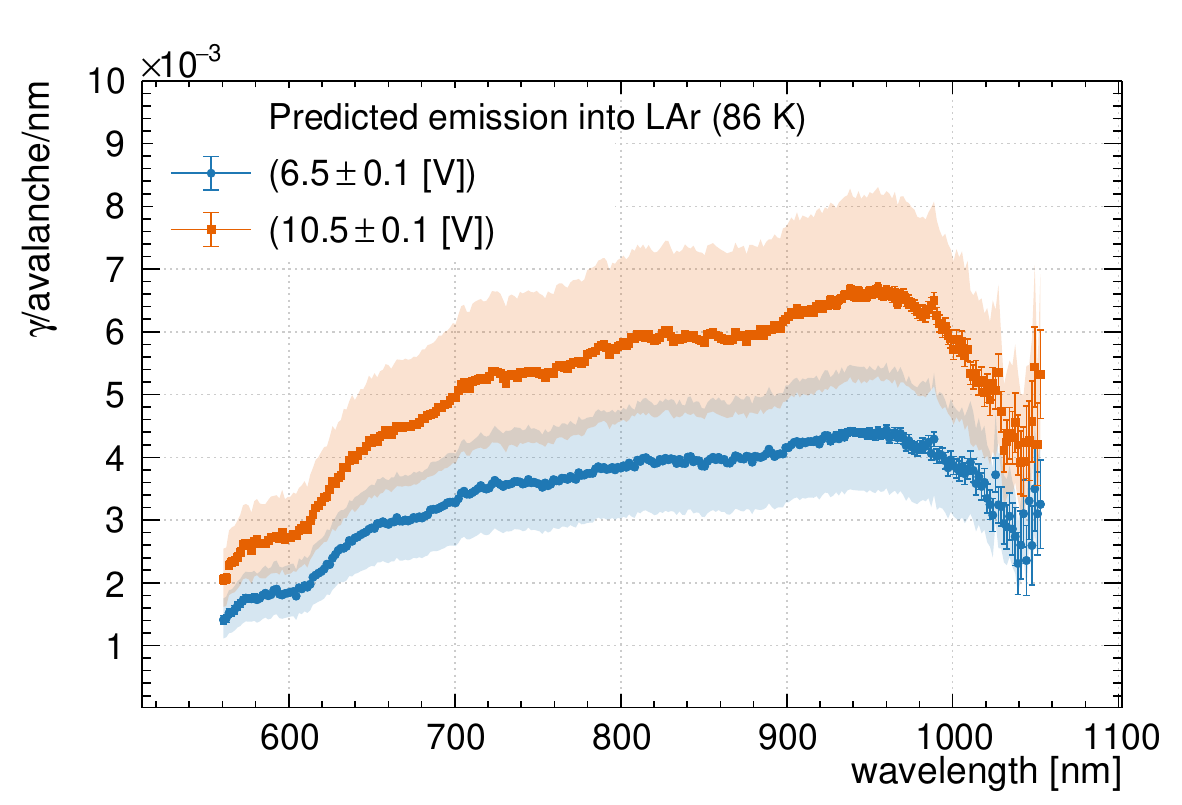}
\caption{Secondary-photon emission spectra $ n_{\gamma}^{\mathrm{LAr}}(\lambda)$ extrapolated to liquid argon at 86~K, shown for over voltages of roughly 6.5 and 10.5~V. The spectra are expressed as photons per avalanche per nanometre.}
\label{fig:eCT_scan_LAr_source}
\end{figure}

Fig.~\ref{fig:eCT_scan_LAr_source} shows the secondary-photon emission spectrum extrapolated to LAr at over voltages of 6.5 and 10.5~V. The shaded bands represent the propagated systematic uncertainties. They do not include the additional, unquantified uncertainty associated with approximating the actual multilayer coating by a single SiO$_2$ layer. Compared with the spectra collected in vacuum, shown in Fig.~\ref{fig:eCT_scan}, the oscillatory structure is less pronounced in LAr. This reduction arises because the refractive index of LAr is better matched to that of the SiO\(_2\) coating than that of vacuum. The resulting decrease in the refractive-index contrast reduces the Fresnel reflection coefficients at the interface and, consequently, the amplitude of the interference pattern.

\begin{figure}[ht]
\centering
\includegraphics[width=0.99\linewidth]{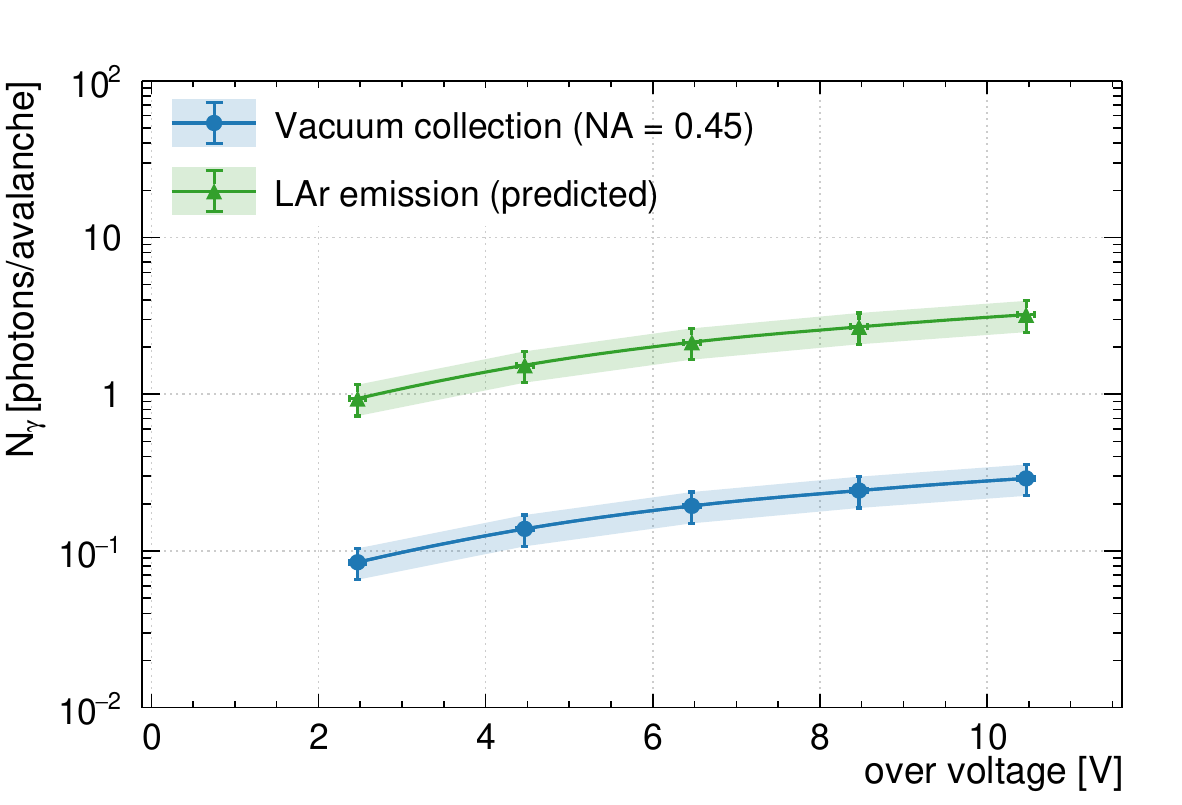}
\caption{Integrated secondary-photon yield ($N_{\gamma}^{m}$, $m\in\{\mathrm{Vac.},\mathrm{LAr}\}$) at 86~K as a function of over voltage. The vacuum data correspond to photons collected within the acceptance of the microscope objective, whereas the LAr results represent the predicted number of photons escaping from the SiPM into the liquid-argon half-space.}
\label{fig:eCT_scan_integrated}
\end{figure}

The spectra shown in Figs.~\ref{fig:eCT_scan} and
\ref{fig:eCT_scan_LAr_source} can be integrated over the measured wavelength range to obtain the corresponding photon yields per avalanche:
\begin{equation}
\label{eq:ngamma}
    N_{\gamma}^{m}
    =
    \int_{550\,\mathrm{nm}}^{1020\,\mathrm{nm}}
    n_{\gamma}^{m}(\lambda)\,d\lambda,
    \qquad
    m\in\{\mathrm{Vac.},\mathrm{LAr}\}.
\end{equation}
Here, \(N_{\gamma}^{\mathrm{Vac.}}\) is the photon yield collected within the acceptance of the microscope objective during the vacuum measurement, whereas \(N_{\gamma}^{\mathrm{LAr}}\) is the predicted yield escaping from the SiPM into the LAr half-space. The latter therefore represents the secondary-photon yield relevant for modeling external optical crosstalk in DS-20k. The integrated yields are shown in Fig.~\ref{fig:eCT_scan_integrated} as a function of over voltage.

The photon yield collected by the microscope objective is substantially lower than the yield predicted to escape into LAr, primarily because the objective subtends only a limited solid angle. In addition, the refractive index of LAr is better matched to those of the silicon dioxide coating and silicon than that of vacuum, reducing internal reflection and increasing the fraction of photons transmitted out of the SiPM.

\section{Conclusions}

This work presents a comprehensive characterization of the FBK NUV-HD-cryo SiPMs selected for the DS-20k experiment. The study covers the main parameters that determine the performance of these devices at cryogenic temperature, including gain, dark-noise rate, prompt and delayed correlated avalanches, photon-detection efficiency, and secondary-photon emission. The electrical and noise properties were characterized as functions of over voltage at 77~K using measurements performed with multiple experimental setups. Where possible, the results were compared with measurements performed by FBK on preliminary engineering and pre-production devices. The photon-detection efficiency was measured as a function of wavelength and over voltage, covering the spectral region relevant to the detection of TPB-shifted liquid-argon scintillation light. The measurements also revealed an oscillatory spectral structure associated with optical interference in the SiPM surface-coating layers.\\The PDE expected in liquid argon was estimated from the vacuum measurements using simulations of the wavelength-dependent ARC transmittance in the two media. The secondary-photon emission produced during SiPM avalanches was characterized at cryogenic temperature over the near-infrared wavelength range. The measured spectra were corrected for the wavelength-dependent efficiency and finite angular acceptance of the experimental apparatus. An optical model was then used to estimate the number and spectral distribution of secondary photons escaping from the SiPM into liquid argon. These measurements provide essential inputs for simulations of external optical crosstalk in DS-20k, where secondary photons can propagate between photosensors and generate additional avalanches. Such additional avalanches can affect pulse identification and bias the reconstructed light yield, potentially degrading the energy resolution. The secondary-emission extrapolation to liquid argon is currently limited by the available knowledge of the SiPM surface structure. In this calculation, the ARC was represented using a simplified single-layer optical model because the composition and thicknesses of its individual layers were not available. A more detailed characterisation of the coating would allow a more realistic model and an assessment of the associated systematic uncertainty. Overall, the measurements presented in this work provide a consistent characterization of the principal SiPM properties required for the simulation and operation of the DS-20k optical system. They also establish a reference dataset for detector-response modelling, SiPM quality assessment, and future studies of large-area photosensor systems operated in liquid argon.

\section{Acknowledgment}
\small{The authors thank P.~Margetak and N.~Massacret (TRIUMF) for their help during the commissioning of the TRIUMF testing setup. The authors gratefully acknowledge support by the U. S. National Science Foundation (NSF) (Grants No. PHY-0919363, No. PHY-1004054, No. PHY-1004072, No. PHY-1242585, No. PHY-1314483, No. PHY-2310091, No. PHY- 1314507,
associated collaborative grants, No. PHY-1211308, No. PHY-1314501, No. PHY-1455351 and No. PHY-1606912, as well as Major Research Instrumentation Grant No. MRI-1429544), the Italian Istituto Nazionale di Fisica Nucleare (Grants from Italian Ministero dell'Istruzione, Universita, e Ricerca Progetto Premiale 2013 and Commissione Scientifica Nazionale II), the Natural Sciences and Engineering Research Council of Canada, SNOLAB, and the Arthur B. McDonald Canadian Astroparticle Physics Research Institute.  We acknowledge the financial support by LabEx UnivEarthS (ANR-10-LABX-0023 and ANR18-IDEX-0001), Chinese Academy of Sciences (113111KYSB20210030) and National Natural Science Foundation of China (12020101004). This work has been supported by the S\~{a}o Paulo Research Foundation (FAPESP) grant 2021/11489-7. I.~Albuquerque is partially supported by Conselho Nacional de Desenvolvimento Cient\'{\i}fico e Tecnol\'ogico (CNPq). The Polish Ministry of Science and Higher Education (MNiSW, grant number 6811/IA/SP/2018), the National Science Centre, Poland (UMO-2022/47/B/ST2/02015 and
2021/42/E/ST2/00331), the International Research Agenda Programmes of the Foundation for
Polish Science: AstroCeNT (MAB/2018/7), funded from the European Regional Development Fund, and Astrocent (FENG.02.01-IP.05-A015/25), co-financed by the European
Union under the European Funds for Smart Economy 2021-2027 (FENG);
Teaming for Excellence grant Astrocent Plus (101137080) funded by the
European Union with complementary national funding from the Polish
Ministry of Science and Higher Education (MNiSW/2025/DIR/811), the Science and Technology Facilities Council, part of the United Kingdom Research and Innovation, The Royal Society (United Kingdom), and IN2P3-COPIN consortium (Grant No. 20-152). We also wish to acknowledge the support from Pacific Northwest National Laboratory, which is operated by Battelle for the U.S. Department of Energy under Contract No. DE-AC05-76RL01830. This research was also supported by the Fermi National Accelerator Laboratory (Fermilab), a U.S. Department of Energy, Office of Science, HEP User Facility. Fermilab is managed by Fermi Research Alliance, LLC (FRA), acting under Contract No. DE-AC02-07CH11359.}

\bibliographystyle{unsrt}
\bibliography{bibliography}

@article{Itzler10102012,
  author = {Mark A. Itzler and Xudong Jiang and Mark Entwistle},
  title = {{Power law temporal dependence of InGaAs/InP SPAD afterpulsing}},
  journal = {Journal of Modern Optics},
  volume = {59},
  number = {17},
  pages = {1472--1480},
  year = {2012},
  doi = {10.1080/09500340.2012.698659},
  note = {DOI: \href{https://doi.org/10.1080/09500340.2012.698659}{\nolinkurl{10.1080/09500340.2012.698659}}},
}

@article{Boulay2021,
  author = {M. G. Boulay and others},
  title = {{Direct comparison of PEN and TPB wavelength shifters in a liquid argon detector}},
  journal = {The European Physical Journal C},
  volume = {81},
  number = {12},
  pages = {1099},
  year = {2021},
  doi = {10.1140/epjc/s10052-021-09870-7},
  note = {DOI: \href{https://doi.org/10.1140/epjc/s10052-021-09870-7}{\nolinkurl{10.1140/epjc/s10052-021-09870-7}}},
  issn = {1434-6044},
}

@article{Ako,
  author = {Jamil, A. and others},
  title = {{VUV-Sensitive Silicon Photomultipliers for Xenon Scintillation Light Detection in nEXO}},
  journal = {IEEE Transactions on Nuclear Science},
  volume = {65},
  number = {11},
  pages = {2823--2833},
  year = {2018},
  doi = {10.1109/TNS.2018.2875668},
  archiveprefix = {arXiv},
  eprint = {1806.02220},
  note = {DOI: \href{https://doi.org/10.1109/TNS.2018.2875668}{\nolinkurl{10.1109/TNS.2018.2875668}}},
  issn = {0018-9499},
}

@article{Aprile2015,
  author = {E. Aprile and others},
  title = {{Lowering the radioactivity of the photomultiplier tubes for the XENON1T dark matter experiment}},
  journal = {The European Physical Journal C},
  volume = {75},
  number = {11},
  pages = {546},
  year = {2015},
  doi = {10.1140/epjc/s10052-015-3657-5},
  url = {http://link.springer.com/10.1140/epjc/s10052-015-3657-5},
  note = {DOI: \href{https://doi.org/10.1140/epjc/s10052-015-3657-5}{\nolinkurl{10.1140/epjc/s10052-015-3657-5}}},
  issn = {1434-6044},
}

@article{Abe2019,
  author = {K. Abe and others},
  title = {{Development of low radioactivity photomultiplier tubes for the XMASS-I detector}},
  journal = {Nuclear Instruments and Methods in Physics Research Section A},
  volume = {922},
  pages = {171--176},
  year = {2019},
  doi = {10.1016/j.nima.2018.12.083},
  url = {https://linkinghub.elsevier.com/retrieve/pii/S0168900218318977},
  note = {DOI: \href{https://doi.org/10.1016/j.nima.2018.12.083}{\nolinkurl{10.1016/j.nima.2018.12.083}}},
  issn = {01689002},
}

@article{Garutti2011,
  author = {Garutti, E.},
  title = {{Silicon photomultipliers for high energy physics detectors}},
  journal = {Journal of Instrumentation},
  volume = {6},
  number = {10},
  pages = {C10003},
  year = {2011},
  doi = {10.1088/1748-0221/6/10/C10003},
  archiveprefix = {arXiv},
  eprint = {1108.3166},
  note = {DOI: \href{https://doi.org/10.1088/1748-0221/6/10/C10003}{\nolinkurl{10.1088/1748-0221/6/10/C10003}}},
  issn = {1748-0221},
}

@article{Aalseth2018,
  author = {Aalseth, C. E. and others},
  title = {{DarkSide-20k: A 20 tonne two-phase LAr TPC for direct dark matter detection at LNGS}},
  journal = {The European Physical Journal Plus},
  volume = {133},
  number = {3},
  pages = {131},
  year = {2018},
  doi = {10.1140/epjp/i2018-11973-4},
  url = {http://link.springer.com/10.1140/epjp/i2018-11973-4},
  archiveprefix = {arXiv},
  eprint = {1707.08145v1},
  note = {DOI: \href{https://doi.org/10.1140/epjp/i2018-11973-4}{\nolinkurl{10.1140/epjp/i2018-11973-4}}},
  issn = {2190-5444},
}

@misc{wayne,
  title = {{Wayne Kerr Electronics}},
  howpublished = {\url{http://www.waynekerrtest.com/index.php}},
}

@article{Hurkx1992,
  author = {Hurkx, G.A.M. and others},
  title = {{A new analytical diode model including tunneling and avalanche breakdown}},
  journal = {IEEE Transactions on Electron Devices},
  volume = {39},
  number = {9},
  pages = {2090--2098},
  year = {1992},
  doi = {10.1109/16.155882},
  url = {http://ieeexplore.ieee.org/document/155882/},
  note = {DOI: \href{https://doi.org/10.1109/16.155882}{\nolinkurl{10.1109/16.155882}}},
  issn = {00189383},
}

@article{Gallina2019,
  author = {Gallina, G. and others},
  title = {{Characterization of the Hamamatsu VUV4 MPPCs for nEXO}},
  journal = {Nuclear Instruments and Methods in Physics Research Section A},
  volume = {940},
  pages = {371--379},
  year = {2019},
  doi = {10.1016/j.nima.2019.05.096},
  archiveprefix = {arXiv},
  eprint = {1903.03663},
  note = {DOI: \href{https://doi.org/10.1016/j.nima.2019.05.096}{\nolinkurl{10.1016/j.nima.2019.05.096}}},
  issn = {01689002},
}

@article{Wagenpfeil2021,
  author = {Wagenpfeil, M. and others},
  title = {{Reflectivity of VUV-sensitive silicon photomultipliers in liquid Xenon}},
  journal = {Journal of Instrumentation},
  volume = {16},
  number = {08},
  pages = {P08002},
  year = {2021},
  doi = {10.1088/1748-0221/16/08/P08002},
  note = {DOI: \href{https://doi.org/10.1088/1748-0221/16/08/P08002}{\nolinkurl{10.1088/1748-0221/16/08/P08002}}},
  issn = {1748-0221},
}

@article{Nakarmi2020,
  author = {Nakarmi, P. and others},
  title = {{Reflectivity and PDE of VUV4 Hamamatsu SiPMs in liquid xenon}},
  journal = {Journal of Instrumentation},
  volume = {15},
  number = {01},
  pages = {P01019},
  year = {2020},
  doi = {10.1088/1748-0221/15/01/P01019},
  url = {https://iopscience.iop.org/article/10.1088/1748-0221/15/01/P01019},
  note = {DOI: \href{https://doi.org/10.1088/1748-0221/15/01/P01019}{\nolinkurl{10.1088/1748-0221/15/01/P01019}}},
  issn = {1748-0221},
}

@article{FALCONE2021164648,
  author = {A. Falcone and others},
  title = {{Cryogenic SiPM arrays for the DUNE photon detection system}},
  journal = {Nuclear Instruments and Methods in Physics Research Section A},
  volume = {985},
  pages = {164648},
  year = {2021},
  doi = {10.1016/j.nima.2020.164648},
  note = {DOI: \href{https://doi.org/10.1016/j.nima.2020.164648}{\nolinkurl{10.1016/j.nima.2020.164648}}},
  issn = {0168-9002},
}

@article{Gallina2019avalanche,
  author = {Gallina, G. and others},
  title = {{Characterization of SiPM Avalanche Triggering Probabilities}},
  journal = {IEEE Transactions on Electron Devices},
  volume = {66},
  number = {10},
  pages = {4228--4234},
  year = {2019},
  doi = {10.1109/TED.2019.2935690},
  note = {DOI: \href{https://doi.org/10.1109/TED.2019.2935690}{\nolinkurl{10.1109/TED.2019.2935690}}},
}

@article{Butcher2017,
  author = {Butcher, A. and others},
  title = {{A method for characterizing after-pulsing and dark noise of PMTs and SiPMs}},
  journal = {Nuclear Instruments and Methods in Physics Research Section A},
  volume = {875},
  pages = {87--91},
  year = {2017},
  doi = {10.1016/j.nima.2017.08.035},
  archiveprefix = {arXiv},
  eprint = {1703.06204},
  note = {DOI: \href{https://doi.org/10.1016/j.nima.2017.08.035}{\nolinkurl{10.1016/j.nima.2017.08.035}}},
  issn = {01689002},
}

@article{Acerbi2017,
  author = {Acerbi, F. and others},
  title = {{Cryogenic Characterization of FBK HD Near-UV Sensitive SiPMs}},
  journal = {IEEE Transactions on Electron Devices},
  volume = {64},
  number = {2},
  pages = {521--526},
  year = {2017},
  doi = {10.1109/TED.2016.2641586},
  url = {http://ieeexplore.ieee.org/document/7807295/},
  note = {DOI: \href{https://doi.org/10.1109/TED.2016.2641586}{\nolinkurl{10.1109/TED.2016.2641586}}},
  issn = {0018-9383},
}

@article{Capasso2020,
  author = {Capasso, M. and others},
  title = {{FBK VUV-sensitive Silicon Photomultipliers for cryogenic temperatures}},
  journal = {Nuclear Instruments and Methods in Physics Research Section A},
  volume = {982},
  pages = {164478},
  year = {2020},
  doi = {10.1016/j.nima.2020.164478},
  note = {DOI: \href{https://doi.org/10.1016/j.nima.2020.164478}{\nolinkurl{10.1016/j.nima.2020.164478}}},
  issn = {01689002},
}

@phdthesis{gallina2021development,
  author = {Gallina, G.},
  title = {{Development of a single vacuum ultra-violet photon-sensing solution for nEXO}},
  school = {University of British Columbia},
  year = {2021},
  doi = {10.14288/1.0396697},
  howpublished = {\url{https://open.library.ubc.ca/collections/ubctheses/24/items/1.0396697}},
  note = {DOI: \href{https://doi.org/10.14288/1.0396697}{\nolinkurl{10.14288/1.0396697}}},
  series = {Electronic Theses and Dissertations (ETDs) 2008+},
}

@article{Boone2017,
  author = {Boone, K. and others},
  title = {{Delayed avalanches in Multi-Pixel Photon Counters}},
  journal = {Journal of Instrumentation},
  volume = {12},
  number = {07},
  pages = {P07026},
  year = {2017},
  doi = {10.1088/1748-0221/12/07/P07026},
  note = {DOI: \href{https://doi.org/10.1088/1748-0221/12/07/P07026}{\nolinkurl{10.1088/1748-0221/12/07/P07026}}},
  issn = {1748-0221},
}

@article{Baldini2018,
  author = {A. M. Baldini and others},
  title = {{The design of the {MEG II} experiment}},
  journal = {European Physical Journal C},
  volume = {78},
  number = {5},
  pages = {380},
  year = {2018},
  doi = {10.1140/epjc/s10052-018-5845-6},
  note = {DOI: \href{https://doi.org/10.1140/epjc/s10052-018-5845-6}{\nolinkurl{10.1140/epjc/s10052-018-5845-6}}},
  issn = {14346052},
}

@article{Acerbi2022,
  author = {F. Acerbi and others},
  title = {{NUV and VUV sensitive Silicon Photomultipliers technologies optimized for operation at cryogenic temperatures}},
  journal = {Nuclear Instruments and Methods in Physics Research Section A},
  volume = {1046},
  pages = {167683},
  year = {2023},
  doi = {10.1016/j.nima.2022.167683},
  note = {DOI: \href{https://doi.org/10.1016/j.nima.2022.167683}{\nolinkurl{10.1016/j.nima.2022.167683}}},
  issn = {01689002},
}

@article{Zhang2022,
  author = {Zhang, J. and others},
  title = {{Scintillation detectors with silicon photomultiplier readout in a dilution refrigerator at temperatures down to 0.2 K}},
  journal = {Journal of Instrumentation},
  volume = {17},
  number = {06},
  pages = {P06024},
  year = {2022},
  doi = {10.1088/1748-0221/17/06/P06024},
  note = {DOI: \href{https://doi.org/10.1088/1748-0221/17/06/P06024}{\nolinkurl{10.1088/1748-0221/17/06/P06024}}},
  issn = {1748-0221},
}

@article{Gola2019,
  author = {A. Gola and others},
  title = {{NUV-Sensitive Silicon Photomultiplier Technologies Developed at Fondazione Bruno Kessler}},
  journal = {Sensors},
  volume = {19},
  number = {2},
  pages = {308},
  year = {2019},
  doi = {10.3390/s19020308},
  url = {http://www.mdpi.com/1424-8220/19/2/308},
  note = {DOI: \href{https://doi.org/10.3390/s19020308}{\nolinkurl{10.3390/s19020308}}},
  issn = {1424-8220},
}

@article{DIncecco2018,
  author = {M. D'Incecco and others},
  title = {{Development of a Novel Single-Channel, 24 $\mathrm{cm}^{2}$, SiPM-Based, Cryogenic Photodetector}},
  journal = {IEEE Transactions on Nuclear Science},
  volume = {65},
  number = {1},
  pages = {591--596},
  year = {2018},
  doi = {10.1109/TNS.2017.2774779},
  url = {http://ieeexplore.ieee.org/document/8113575/},
  note = {DOI: \href{https://doi.org/10.1109/TNS.2017.2774779}{\nolinkurl{10.1109/TNS.2017.2774779}}},
  issn = {0018-9499},
}

@article{Francini2013,
  author = {R. Francini and others},
  title = {{VUV-Vis optical characterization of Tetraphenyl-butadiene films on glass and specular reflector substrates from room to liquid Argon temperature}},
  journal = {Journal of Instrumentation},
  volume = {8},
  number = {09},
  pages = {P09006},
  year = {2013},
  doi = {10.1088/1748-0221/8/09/P09006},
  note = {DOI: \href{https://doi.org/10.1088/1748-0221/8/09/P09006}{\nolinkurl{10.1088/1748-0221/8/09/P09006}}},
  issn = {1748-0221},
}

@article{mirzoyan2009light,
  author = {R. Mirzoyan and others},
  title = {{Light emission in Si avalanches}},
  journal = {Nuclear Instruments and Methods in Physics Research Section A},
  volume = {610},
  number = {1},
  pages = {98--100},
  year = {2009},
  doi = {10.1016/j.nima.2009.05.081},
  url = {https://linkinghub.elsevier.com/retrieve/pii/S0168900209010377},
  note = {DOI: \href{https://doi.org/10.1016/j.nima.2009.05.081}{\nolinkurl{10.1016/j.nima.2009.05.081}}},
  issn = {01689002},
}

@article{instruments5010004,
  author = {Ku\'zniak, M. and Szelc, A. M.},
  title = {{Wavelength Shifters for Applications in Liquid Argon Detectors}},
  journal = {Instruments},
  volume = {5},
  number = {1},
  pages = {4},
  year = {2021},
  doi = {10.3390/instruments5010004},
  url = {https://www.mdpi.com/2410-390X/5/1/4},
  note = {DOI: \href{https://doi.org/10.3390/instruments5010004}{\nolinkurl{10.3390/instruments5010004}}},
  issn = {2410-390X},
}

@article{piemonte2016performance,
  author = {Piemonte, C. and others},
  title = {{Performance of NUV-HD silicon photomultiplier technology}},
  journal = {IEEE Transactions on Electron Devices},
  volume = {63},
  number = {3},
  pages = {1111--1116},
  year = {2016},
}

@article{ORGANTINI2020164410,
  author = {P.~Organtini and others},
  title = {{Industrial exploitation of SiPM technology developed for basic research}},
  journal = {Nuclear Instruments and Methods in Physics Research Section A},
  volume = {978},
  pages = {164410},
  year = {2020},
  doi = {10.1016/j.nima.2020.164410},
  note = {DOI: \href{https://doi.org/10.1016/j.nima.2020.164410}{\nolinkurl{10.1016/j.nima.2020.164410}}},
  issn = {0168-9002},
}

@article{20kcollaboration2024qaqcsipm,
  author = {F. Acerbi and others},
  title = {{Quality Assurance and Quality Control of the $26\,\mathrm{m}^{2}$ SiPM production for the DarkSide-20k dark matter experiment}},
  journal = {The European Physical Journal C},
  volume = {85},
  pages = {534},
  year = {2025},
  doi = {10.1140/epjc/s10052-025-14196-9},
  archiveprefix = {arXiv},
  eprint = {2412.18867},
  note = {DOI: \href{https://doi.org/10.1140/epjc/s10052-025-14196-9}{\nolinkurl{10.1140/epjc/s10052-025-14196-9}}},
}

@article{Kugathasan:2020xry,
  author = {Kugathasan, R.},
  title = {{A low-power mixed-signal ASIC for readout of SiPM at cryogenic temperature}},
  journal = {PoS},
  volume = {TWEPP2019},
  pages = {011},
  year = {2020},
  doi = {10.22323/1.370.0011},
  note = {DOI: \href{https://doi.org/10.22323/1.370.0011}{\nolinkurl{10.22323/1.370.0011}}},
}

@article{Gallina2022,
  author = {G. Gallina and others},
  title = {{Performance of novel VUV-sensitive Silicon Photo-Multipliers for nEXO}},
  journal = {The European Physical Journal C},
  volume = {82},
  number = {12},
  pages = {1125},
  year = {2022},
  doi = {10.1140/epjc/s10052-022-11072-8},
  url = {https://link.springer.com/10.1140/epjc/s10052-022-11072-8},
  note = {DOI: \href{https://doi.org/10.1140/epjc/s10052-022-11072-8}{\nolinkurl{10.1140/epjc/s10052-022-11072-8}}},
  issn = {1434-6052},
}

@article{RazetoElectronics2018,
  author = {M. D'Incecco and others},
  title = {{Development of a Very Low-Noise Cryogenic Preamplifier for Large-Area SiPM Devices}},
  journal = {IEEE Transactions on Nuclear Science},
  volume = {65},
  number = {4},
  pages = {1005--1011},
  year = {2018},
  doi = {10.1109/TNS.2018.2799325},
  note = {DOI: \href{https://doi.org/10.1109/TNS.2018.2799325}{\nolinkurl{10.1109/TNS.2018.2799325}}},
  issn = {0018-9499},
}

@article{Razeto2023,
  author = {Boulay, M. G. and others},
  title = {{SiPM cross-talk in liquid argon detectors}},
  journal = {Frontiers in Physics},
  volume = {11},
  pages = {1181400},
  year = {2023},
  doi = {10.3389/fphy.2023.1181400},
  note = {DOI: \href{https://doi.org/10.3389/fphy.2023.1181400}{\nolinkurl{10.3389/fphy.2023.1181400}}},
  issn = {2296-424X},
}

@article{10713286,
  author = {Raymond, K. and others},
  title = {{Stimulated Secondary Emission of Single-Photon Avalanche Diodes}},
  journal = {IEEE Transactions on Electron Devices},
  volume = {71},
  number = {11},
  pages = {6871--6879},
  year = {2024},
  doi = {10.1109/TED.2024.3469918},
  note = {DOI: \href{https://doi.org/10.1109/TED.2024.3469918}{\nolinkurl{10.1109/TED.2024.3469918}}},
}

@article{GRACE2017204,
  author = {E. Grace and others},
  title = {{Index of refraction, Rayleigh scattering length, and Sellmeier coefficients in solid and liquid argon and xenon}},
  journal = {Nuclear Instruments and Methods in Physics Research Section A},
  volume = {867},
  pages = {204--208},
  year = {2017},
  doi = {10.1016/j.nima.2017.06.031},
  url = {https://www.sciencedirect.com/science/article/pii/S0168900217306848},
  note = {DOI: \href{https://doi.org/10.1016/j.nima.2017.06.031}{\nolinkurl{10.1016/j.nima.2017.06.031}}},
  issn = {0168-9002},
}

@article{VINOGRADOV2012247,
  author = {S. Vinogradov},
  title = {{Analytical models of probability distribution and excess noise factor of solid state photomultiplier signals with crosstalk}},
  journal = {Nuclear Instruments and Methods in Physics Research Section A},
  volume = {695},
  pages = {247--251},
  year = {2012},
  doi = {10.1016/j.nima.2011.11.086},
  url = {https://www.sciencedirect.com/science/article/pii/S0168900211021565},
  note = {New Developments in Photodetection NDIP11; DOI: \href{https://doi.org/10.1016/j.nima.2011.11.086}{\nolinkurl{10.1016/j.nima.2011.11.086}}},
  issn = {0168-9002},
}

@article{veto20k,
  author = {Acerbi, F. and others},
  title = {{Construction and characterisation of the DarkSide-20k veto silicon photo-multiplier tiles}},
  journal = {The European Physical Journal C},
  volume = {86},
  pages = {1089},
  year = {2026},
  doi = {10.1140/epjc/s10052-026-16217-7},
  note = {DOI: \href{https://doi.org/10.1140/epjc/s10052-026-16217-7}{\nolinkurl{10.1140/epjc/s10052-026-16217-7}}},
}

@article{Aprile_2022,
  author = {Aprile, E. and others},
  title = {{Material radiopurity control in the XENONnT experiment}},
  journal = {The European Physical Journal C},
  volume = {82},
  number = {7},
  pages = {599},
  year = {2022},
  doi = {10.1140/epjc/s10052-022-10345-6},
  note = {DOI: \href{https://doi.org/10.1140/epjc/s10052-022-10345-6}{\nolinkurl{10.1140/epjc/s10052-022-10345-6}}},
  issn = {1434-6052},
}

@article{Agnes2024UAr,
  author = {Agnes, P. and others},
  title = {{The underground argon project: procurement and purification of argon for
                   dark matter searches and beyond}},
  journal = {Front. Phys.},
  volume = {12},
  pages = {1387069},
  year = {2024},
  doi = {10.3389/fphy.2024.1387069},
  note = {DOI: \href{https://doi.org/10.3389/fphy.2024.1387069}{\nolinkurl{10.3389/fphy.2024.1387069}}},
  ids = {10.3389/fphy.2024.1387069},
}

@article{DUNE1,
  author = {{DUNE Collaboration}},
  title = {{Long-baseline neutrino oscillation physics potential of the DUNE
                   experiment}},
  journal = {Eur. Phys. J. C},
  volume = {80},
  number = {10},
  pages = {978},
  year = {2020},
  doi = {10.1140/epjc/s10052-020-08456-z},
  note = {DOI: \href{https://doi.org/10.1140/epjc/s10052-020-08456-z}{\nolinkurl{10.1140/epjc/s10052-020-08456-z}}},
}

@article{Acerbi_2026,
  author = {Acerbi, F. and others},
  title = {{The DarkSide-20k Data Acquisition System}},
  journal = {Journal of Instrumentation},
  volume = {21},
  number = {09},
  pages = {P09017},
  year = {2026},
  doi = {10.1088/1748-0221/21/09/P09017},
  note = {DOI: \href{https://doi.org/10.1088/1748-0221/21/09/P09017}{\nolinkurl{10.1088/1748-0221/21/09/P09017}}},
}

\newcommand{\notds}{\nolinebreak\footnotemark\nolinebreak}
\renewcommand{\thefootnote}{$*$}
{
\onecolumn
\textbf{The DarkSide-20k Collaboration}

F.~Acerbi\thanksref{l_TNFBK}\nolinebreak,
P.~Adhikari\thanksref{l_Carleton}\nolinebreak,
P.~Agnes\thanksref{l_AQGSSI}\textsuperscript{,}\thanksref{l_AQLNGS}\nolinebreak,
I.~Ahmad\thanksref{l_UCRiverside}\nolinebreak,
S.~Albergo\thanksref{l_CTINFN}\textsuperscript{,}\thanksref{l_CTUNI}\nolinebreak,
I.~F.~Albuquerque\thanksref{l_USP}\nolinebreak,
T.~Alexander\thanksref{l_PNNL}\nolinebreak,
A.~K.~Alton\thanksref{l_Augustana}\nolinebreak,
P.~Amaudruz\thanksref{l_TRIUMF}\nolinebreak,
G.~A.~Anastasi\thanksref{l_CTINFN}\textsuperscript{,}\thanksref{l_CTUNI}\nolinebreak
% ...

\begin{enumerate}
\item{Fondazione Bruno Kessler, Povo, 38123, Italy\label{l_TNFBK}}
\item{Department of Physics, Carleton University, Ottawa, ON K1S 5B6, Canada\label{l_Carleton}}
\item{Gran Sasso Science Institute, L'Aquila, 67100, Italy\label{l_AQGSSI}}
\item{INFN Laboratori Nazionali del Gran Sasso, Assergi (AQ), 67100, Italy\label{l_AQLNGS}}
\item{Center for Experimental Cosmology \& Instrumentation, Department of Physics and Astronomy, University of California, Riverside, CA 92507, USA\label{l_UCRiverside}}
\item{INFN Catania, Catania, 95121, Italy\label{l_CTINFN}}
\item{Universit\`a of Catania, Catania, 95124, Italy\label{l_CTUNI}}
\item{Instituto de F\'isica, Universidade de S\~ao Paulo, S\~ao Paulo, 05508-090, Brazil\label{l_USP}}
\item{Pacific Northwest National Laboratory, Richland, WA 99352, USA\label{l_PNNL}}
\item{Physics Department, Augustana University, Sioux Falls, SD 57197, USA\label{l_Augustana}}
\item{TRIUMF, 4004 Wesbrook Mall, Vancouver, BC V6T 2A3, Canada\label{l_TRIUMF}}
\item{Physics Department, Columbia University, New York, NY 10027, USA\label{l_Columbia}}
\item{Department of Physics, University of Alberta, Edmonton, AB T6G 2R3, Canada\label{l_Alberta}}
\item{INFN Laboratori Nazionali di Legnaro, Legnaro (Padova), 35020, Italy\label{l_LNLINFN}}
\item{Savannah River National Laboratory, Aiken, SC 29808, United States\label{l_SNL}}
\item{Institute for Experimental Physics, University of Hamburg, Hamburg, 22761, Germany\label{l_UniHAM}}
\item{CIEMAT, Centro de Investigaciones Energ\'eticas, Medioambientales y Tecnol\'ogicas, Madrid, 28040, Spain\label{l_CIEMAT}}
\item{Centre de Physique des Particules de Marseille, Aix Marseille Univ, CNRS/IN2P3, CPPM, Marseille, France\label{l_CPPM}}
\item{INFN Pisa, Pisa, 56127, Italy\label{l_PIINFN}}
\item{Physics Department, Universit\`a degli Studi di Pisa, Pisa, 56127, Italy\label{l_PIUniPHY}}
\item{Centro de Astropart\'iculas y F\'isica de Altas Energ\'ias, Universidad de Zaragoza, Zaragoza, 50009, Spain\label{l_Zaragoza}}
\item{ORCID 0000-0002-1767-1754, ORCID 0000-0002-4351-2255, ORCID 0000-0002-6394-9219, ORCID 0000-0002-8468-9540, ORCID 0000-0003-2869-2363, ORCID 0009-0009-0770-8830, ORCID 0000-0002-0597-2234, ORCID 0000-0001-5532-7711, ORCID 0000-0002-1455-4341, ORCID 0000-0003-2141-8230, ORCID 0009-0005-0286-0156, ORCID 0000-0001-5128-9279, ORCID 0000-0002-5527-4880, ORCID 0000-0002-5238-0442, ORCID 0000-0003-2952-6412, ORCID 0000-0002-2624-9416\label{l_NewAuthorsORCID}}
\item{Department of Physics, University of Oxford, Oxford, OX1 3RH, UK\label{l_Oxford}}
\item{INFN Torino, Torino, 10125, Italy\label{l_TOINFN}}
\item{Department of Electronics and Communications, Politecnico di Torino, Torino, 10129, Italy\label{l_TOPoli}}
\item{INFN Sezione di Roma, Roma, 00185, Italy\label{l_RMUnoINFN}}
\item{INFN Cagliari, Cagliari, 09042, Italy\label{l_CAINFN}}
\item{Physics Department, Universit\`a degli Studi di Genova, Genova, 16146, Italy\label{l_GEUni}}
\item{INFN Genova, Genova, 16146, Italy\label{l_GEINFN}}
\item{Warsaw University of Technology, Warsaw, 00-661, Poland\label{l_WUT}}
\item{INFN Napoli, Napoli, 80126, Italy\label{l_NAINFN}}
\item{Physics Department, Universit\`a degli Studi ``Federico II'' di Napoli, Napoli, 80126, Italy\label{l_NAUniPHY}}
\item{Department of Physics and Astronomy, Universit\`a degli Studi di Bologna, Bologna, 40126, Italy\label{l_BOUniPHY}}
\item{INFN Bologna, Bologna, 40126, Italy\label{l_BOINFN}}
\item{Virginia Tech, Blacksburg, VA 24061, USA\label{l_VTech}}
\item{AstroCeNT, Nicolaus Copernicus Astronomical Center of the Polish Academy of Sciences, Warsaw, 00-614, Poland\label{l_AstroCeNT}}
\item{M. Smoluchowski Institute of Physics, Jagiellonian University, Krakow, 30-348, Poland\label{l_Krakow}}
\item{Physics Department, Universit\`a degli Studi di Milano, Milano, 20133, Italy\label{l_MIUni}}
\item{INFN Milano, Milano, 20133, Italy\label{l_MIINFN}}
\item{Science \& Technology Facilities Council (STFC), Rutherford Appleton Laboratory, Technology, Harwell Oxford, Didcot, OX11 0QX, UK\label{l_STFCInterconnect}}
\item{Physics Department, Sapienza Universit\`a di Roma, Roma, 00185, Italy\label{l_RMUnoUni}}
\item{Department of Physics, Engineering Physics and Astronomy, Queen's University, Kingston, ON K7L 3N6, Canada\label{l_Queens}}
\item{Department of Physics, University of California Davis, Davis, CA 95616, USA\label{l_UCDavis}}
\item{School of Physics and Astronomy, University of Edinburgh, Edinburgh, EH9 3FD, UK\label{l_UniversityofEdinburgh}}
\item{Physics Department, Universit\`a degli Studi di Cagliari, Cagliari, 09042, Italy\label{l_CAUniPHY}}
\item{APC, Universit\'e de Paris Cit\'e, CNRS, Astroparticule et Cosmologie, Paris, F-75013, France\label{l_APC}}
\item{Department of Physics, University of Liverpool, The Oliver Lodge Laboratory, Liverpool, L69 7ZE, UK\label{l_Liverpool}}
\item{Physics Department, Princeton University, Princeton, NJ 08544, USA\label{l_Princeton}}
\item{Museo Storico della Fisica e Centro Studi e Ricerche Enrico Fermi, Roma, 00184, Italy\label{l_BOCentroFermi}}
\item{Department of Physics and Astronomy, The University of Manchester, Manchester, M13 9PL, UK\label{l_Manchester}}
\item{Williams College, Department of Physics and Astronomy, Williamstown, MA 01267, USA\label{l_WilliamsCollege}}
\item{Department of Physics, Enrico Fermi Institute and Kavli Institute for Cosmological Physics, The University of Chicago, Chicago, IL 60637, USA\label{l_Chicago}}
\item{INFN Laboratori Nazionali del Sud, Catania, 95123, Italy\label{l_CTLNS}}
\item{Engineering and Architecture Department, Universit\`a di Enna Kore, Enna, 94100, Italy\label{l_ENUniCEE}}
\item{Department of Physics and Astronomy, University of Hawai'i, Honolulu, HI 96822, USA\label{l_Hawaii}}
\item{Department of Physics, University of Houston, Houston, TX 77204, USA\label{l_Houston}}
\item{Department of Physics and Astronomy, Laurentian University, Sudbury, ON P3E 2C6, Canada\label{l_Laurentian}}
\item{SNOLAB, Lively, ON P3Y 1N2, Canada\label{l_SNOLAB}}
\item{Department of Physics, Royal Holloway University of London, Egham, TW20 0EX, UK\label{l_RHUL}}
\item{Pharmacy Department, Universit\`a degli Studi ``Federico II'' di Napoli, Napoli, 80131, Italy\label{l_NAUniPHARM}}
\item{Universit\`a degli Studi dell'Aquila, L'Aquila, 67100, Italy\label{l_UnivAQ}}
\item{INFN Roma Tre, Roma, 00146, Italy\label{l_RMTreINFN}}
\item{Mathematics and Physics Department, Universit\`a degli Studi Roma Tre, Roma, 00146, Italy\label{l_RMTreUni}}
\item{University of Warwick, Department of Physics, Coventry, CV47AL, UK\label{l_Warwick}}
\item{Amherst Center for Fundamental Interactions and Physics Department, University of Massachusetts, Amherst, MA 01003, USA\label{l_UMass}}
\item{Department of Electrical and Electronic Engineering, Universit\`a degli Studi di Cagliari, Cagliari, 09123, Italy\label{l_CAUniEEE}}
\item{Istituto Nazionale di Fisica Nucleare, Roma, 00186, Italia\label{l_INFN}}
\item{School of Physics and Astronomy, University of Birmingham, Edgbaston, Birmingham, B15 2TT, UK\label{l_Birmingham}}
\item{Physics Department, Lancaster University, Lancaster, LA1 4YB, UK\label{l_Lancaster}}
\item{Center for Experimental Nuclear Physics and Astrophysics, and Department of Physics, University of Washington, Seattle, WA 98195, USA\label{l_Seattle}}
\item{Chemical, Materials, and Industrial Production Engineering Department, Universit\`a degli Studi ``Federico II'' di Napoli, Napoli, 80126, Italy\label{l_NAUniCHE}}
\item{Institute of High Energy Physics, Beijing, 100049, China\label{l_IHEP}}
\item{University of Chinese Academy of Sciences, Beijing, 100049, China\label{l_UCAS}}
\item{Department of Physics and Engineering, Fort Lewis College, Durango, CO 81301, USA\label{l_FortLewis}}
\item{Institute of Applied Radiation Chemistry, Lodz University of Technology, Lodz, 93-590, Poland\label{l_Lodz}}
\end{enumerate}
\footnotetext{External collaborator.}
}

\end{document}